\documentclass[12pt,a4paper]{article} 

\usepackage{epsfig}
\usepackage{amssymb}
\usepackage{graphicx}
\usepackage{amsmath}

\def\reff#1{(\ref{#1})}

\newcommand{\1}{1\!\!\!\bot}

\newcommand{\be}{\begin{equation}}
\newcommand{\en}{\end{equation}}
\newcommand{\ee}{\end{equation}}
\newcommand{\bea}{\begin{eqnarray}}
\newcommand{\ena}{\end{eqnarray}}

\newcommand{\tr}{\hbox{tr}_c}
\newcommand{\trc}{\hbox{tr}_c}
\newcommand{\Tr}{\hbox{tr}_c}
\newcommand{\hbo}{\hbox to 1 true cm {\hfill } }

\begin{document}

\vglue 1 truecm

\vbox{ UNITU-THEP-018/2003
\hfill \today
}
  
\vfil
\centerline{\large\bf Propagators and running coupling  from $SU(2)$ }

\centerline{\large\bf lattice gauge theory}

\bigskip
\centerline{J.~C.~R.~Bloch $^a$, A.~Cucchieri $^b$, K.~Langfeld $^c$, 
T.~Mendes $^b$} 
\vspace{1 true cm} 

\centerline{$^a$ DFG Research Center ``Mathematics for Key Technologies''}
\centerline{c/o Weierstrass Institute for Applied Analysis and Stochastics}
\centerline{Mohrenstrasse 39, D-10117 Berlin, Germany}

\medskip 
\medskip 

\centerline{$^b$ Instituto de F\'\i sica de S\~ao Carlos,
                 Universidade de S\~ao Paulo}
\centerline{C.P.\ 369, 13560-970 S\~ao Carlos, SP, Brazil}

\medskip 
\medskip 

\centerline{$^c$ Institut f\"ur Theoretische Physik, Universit\"at 
                 T\"ubingen}
\centerline{D--72076 T\"ubingen, Germany}
  
\vfil
\begin{abstract}

We perform numerical studies of the running coupling constant 
$\alpha_R(p^2)$ and of the gluon and ghost propagators for 
pure $SU(2)$ lattice gauge theory in the minimal Landau gauge.
Different definitions of the gauge fields and 
different gauge-fixing procedures are used respectively for gaining
better control over the approach to the continuum limit and for 
a better understanding of Gribov-copy effects.
We find that the ghost-ghost-gluon-vertex renormalization 
constant is finite in the continuum limit,
confirming earlier results by all-order perturbation theory. 
In the low momentum regime, the gluon form factor is suppressed 
while the ghost form factor is divergent. Correspondingly, the ghost
propagator diverges faster than $1/p^2$
and the gluon propagator appears to be finite.
Precision data for 
the running coupling $\alpha_R (p^2)$ are obtained. These data 
are consistent with an IR fixed point given by
$\lim _{p \to 0} \alpha_R (p^2) = 5(1)$. 

\end{abstract}

\vfil
\hrule width 5truecm
\vskip .2truecm
\begin{quote} 

{\it keywords: gluon propagator, ghost propagator, running coupling 
   constant, $SU(2)$ lattice gauge theory}

\end{quote}
\eject
%


\section{Introduction}

The non-perturbative study of non-Abelian gauge theories
is of great importance for the determination of infrared (IR)
properties such as color confinement and hadronization.
These properties are encoded in the low momentum behavior of 
Yang-Mills Green's functions. Derived from these Green functions, the 
so-called (renormalized) running coupling constant ``$\alpha_R(p^2)$'' plays 
an important role for phenomenological studies and model building. 
While at large momentum the running coupling decreases logarithmically 
with momentum, it rapidly rises at the hadronic energy scale of 
several hundred MeVs thus signaling the breakdown of the 
perturbative approach. Nonperturbative studies of $\alpha_R(p^2)$
may be carried out analytically using the
Dyson-Schwinger equations (DSEs) and numerically through
lattice simulations.

\vskip 0.3cm 
Whereas the high momentum behavior of the running coupling 
is uniquely determined and provided by perturbation theory, 
several definitions of the running coupling in the low 
momentum regime are possible. All of them match with the perturbative 
result at high energies. For example, the corrections to the 
Coulomb law of the static-quark potential may be used to 
define the running coupling~\cite{Michael:1992nj,Bali:1992ru}. 
Alternatively, the finite-size scaling has its imprint on the 
running coupling and may be used for a high-precision measurement 
of the coupling~\cite{Luscher:1993gh,deDivitiis:1993hj}. The
approach adopted in~\cite{Parrinello:wd}--\cite{Furui:2000mq}
is based on extracting the running coupling directly from a vertex 
function. For a recent review of lattice 
calculations for $\alpha_R(p^2)$ see \cite{Cucchieri:2002py}.

\vskip 0.3cm 
In order to obtain Green's functions for the fundamental degrees of 
freedom, gluons and quarks, gauge fixing is necessary.
Despite being gauge dependent, these Green functions play an important 
role for the phenomenological approach to hadron physics. 
Landau gauge is a convenient choice for gauge fixing for 
several reasons: First of all, it is a Lorenz-covariant gauge 
implying that 2-point functions only depend on the square 
of the momentum transfer. Secondly, the renormalization procedure 
is simplified since the ghost-ghost-gluon-vertex renormalization 
constant $\widetilde{Z}_1$ is finite,
at least to all orders of perturbation theory. 
This result --- obtained by Taylor \cite{Taylor:ff} --- is a
particular feature of Landau gauge and allows another definition 
of the running coupling constant, which only requires the 
calculation of 2-point functions: Let $F_R(p^2, \mu ^2)$ and $J_R(p^2,\mu ^2)$ 
denote the form factors (for a renormalization point $\mu $) 
of the gluon and the ghost propagator, respectively; 
the running coupling is then defined  by 
\be
\alpha_R(p^2) \; = \; 
\alpha_R(\mu^2) \; F_R(p^2,\mu ^2) \; J_R^2(p^2, \mu^2) \; . 
\en
(see Subsection \ref{sec:runcoup} below). 

\vskip 0.3cm 
Due to its usefulness for the description of the physics of hadrons, 
the non-per\-turba\-tive approach to low-energy Yang-Mills theory 
by means of the DSEs has attracted 
much interest over the last decade~\cite{Roberts:2000aa,Alkofer:2000wg}. 
The coupled set of continuum DSEs for the renormalized gluon and ghost
propagators in Landau gauge has been recently studied by several groups
\cite{vonSmekal:1997is}--\cite{Bloch:2003yu}.
In all cases it was found that the gluon and ghost form factors 
satisfy simple scaling laws in the IR momentum range 
$p \ll  1 \, \mathrm{GeV} $
\be 
F_R(p^2, \mu^2) \; \propto \left[ p^2 \right]^{\alpha} \; , \hbo 
J_R(p^2, \mu^2) \; \propto \left[ p^2 \right]^{\beta} \;
\label{eq:DSEpred}  
\en 
where the remarkable sum rule holds for the IR exponents $\alpha$ 
and $\beta$: 
\be 
\alpha \; + \; 2 \, \beta \; = \; 0 \, . 
\label{eq:sumrule}  
\en 
It is interesting that this result is rather independent of the
truncation scheme under consideration.
These exponents may be determined from lattice
simulations, assuming the parameterization $\alpha = 2 \kappa$
and $\beta = - \kappa $. Note that for $\kappa > 0$ this
implies a divergent ghost form factor $ J_R(p^2, \mu^2) $ in the IR limit
and a vanishing gluon form factor $ F_R(p^2, \mu^2) $ in the same limit. Also,
since the gluon propagator is given by $D(p^2) = F(p^2) / p^2$, one gets
that $D(0)$ is infinite or finite respectively if $\kappa < 0.5$ or
$\kappa \geq 0.5$. In the second case one has $D(0) = 0$ for
$\kappa > 0.5$ and $D(0)$ finite and non-zero for $\kappa = 0.5$.
The IR sum rule \reff{eq:sumrule} also implies that the running
coupling [defined in eq.\ \reff{eq:alphaRfin} below] 
develops a fixed point in the IR limit 
\be 
\lim_{p\to 0} \alpha_R(p^2) \; = \;
\alpha_c \;=\;\hbox{constant}. 
\label{eq:alphac}
\en 
Note that this result is independent of the value of $\kappa$
as long as the IR sum rule \reff{eq:sumrule} is satisfied. 

\vskip 3mm
The precise value of $ \kappa $ as well as the fixed-point value
$\alpha_c$ depend strongly on the truncation of the Dyson-Schwinger
tower of equations. In fact, depending on the truncation, 
one finds $0.3 < \kappa < 1 $ in the four-dimensional case
\cite{vonSmekal:1997is}--\cite{Bloch:2003yu}.
These studies vary in their vertex Ans\"atze, angular approximations
of the momentum loop integral, and on the tensor structure considered. 
In Ref.\ \cite{Bloch:2001wz} 
a new class of truncation schemes has been introduced, which
manifestly ensures the multiplicative renormalizability 
of the propagator solutions. In this truncation, 
although the exact values of $\kappa$ and $\alpha_c$
depend on the details of the truncation of the DSE tower, the value
of $\alpha_c$ is constrained to  
\be
2\pi/N_c < \alpha_c < 8\pi/N_c
\label{alpha_bounds}
\en
for SU($N_c$). 


\vskip 3mm
An IR-finite gluon propagator
\cite{Cucchieri:1997fy}--\cite{Cucchieri:1997dx}
and an IR-divergent ghost form factor \cite{Suman:1995zg,Cucchieri:1997dx}
are also obtained using numerical simulations in the minimal Landau gauge.
The present numerical data for the ghost propagator indicate a
value of $\kappa = - \beta$ smaller than $0.5$, while for the
gluon propagator it is still under debate if $\kappa = \alpha / 2$ is equal to
or larger than $0.5$. In both cases large finite-size effects in the IR region
make an exact determination of these exponents difficult. This is
particularly evident in the gluon propagator case 
\cite{Cucchieri:1997fy,Cucchieri:1999sz,Bonnet:2001uh,
Cucchieri:2000kw,Cucchieri:2003di}, where one needs to go to very large
lattices in order to have control over the infinite-volume extrapolation.
Our present data are consistent with $\kappa$ of the order of 0.5
\cite{Bloch:2002we,Langfeld:2002dd}.


\vskip 3mm
Let us stress that in the minimal Landau gauge, which is the
gauge-fixing condition used in numerical simulations
(see Section \ref{gaugefix} below), the gauge-fixed
configurations belong to the region of
transverse configurations, for which the Faddeev-Popov operator
is non-negative. This implies a rigorous inequality
\cite{Zwanziger:1990by}--\cite{Zwanziger:1993dh}
for the Fourier components of the gluon field
and a strong suppression of the (unrenormalized)
gluon propagator in the IR limit. At the same time,
the Euclidean probability gets concentrated near
the border of this region, the so-called {\it first Gribov horizon},
implying the enhancement of the ghost propagator at small
momenta \cite{Zwanziger:1993dh}.
A similar result was also obtained by Gribov in \cite{Gribov:1977wm}.


\vskip 3mm
Taylor's finding is based upon the Faddeev-Popov quantization, thereby 
ignoring the effect of Gribov copies, which are certainly present 
in an intrinsically non-perturbative approach. 
The goal of the present paper is to confirm Taylor's result
($\widetilde{Z}_1$ is finite to all orders of perturbation theory) 
by the non-perturbative approach provided by lattice simulations. 
In addition, a thorough study of the gluon and 
the ghost form factors is performed. A focal point is 
the IR limit of the running coupling constant. Support for 
the existence of the fixed point is found, and a first estimate of 
$\alpha _c$ is provided from extensive lattice simulations. 
We present two sets of simulations, carried out respectively in
S\~ao Carlos and in T\"ubingen, employing different definitions of the
gauge fields and different gauge-fixing procedures. We believe that
the comparison of these two formulations strengthens the importance
of our findings.

\vskip 3mm
The paper is organized as follows. In Section \ref{numerical}
we describe the lattice approach to Yang-Mills Green's functions. 
Section \ref{sec:details} contains the numerical setup for 
the simulations carried out in
S\~ao Carlos and in T\"ubingen. In Section \ref{results}
we report our data for the gluon and ghost
propagators and for the running coupling constant. 
Conclusions are left to the final section. 

Preliminary results have been presented in 
\cite{Bloch:2002we,Langfeld:2002dd,Cucchieri:2001za,Langfeld:2002bg}.


\section{The lattice approach to Green's functions} 
\label{numerical}

In this section we explain the two lattice setups used
for the numerical evaluation of the gluon and ghost
propagators. We also recall the definition of the running
coupling constant considered in Ref.~\cite{vonSmekal:1997is},
which can been evaluated using these propagators.


\subsection{Gluon field on the lattice}
\label{sec:gluonfield}

The action $S$ of the continuum $SU(2)$ Yang-Mills theory
is formulated in terms of the field strength
\be
F^{a}_{\mu \nu}[A](x) \;=\; \partial_{\mu} A^{a}_{\nu}(x) \, - \,
\partial_{\nu} A^{a}_{\mu}(x)  \; + \; g_0 \; \epsilon^{abc}
 A^{b}_{\mu}(x) \, A^{c}_{\nu}(x)
\en
and is given by
\be
S \;=\; \frac{1}{4} \int d^{4}x \;
    F^{a}_{\mu \nu}[A](x) \; F^{a}_{\mu \nu}[A](x)
\; .
\label{action}
\en
Here $ g_0 $ is the bare coupling constant and $ A^{a}_{\mu}(x) $ is
the continuum gauge field. 

On the lattice the dynamical fields are $SU(2)$ matrices 
$U_\mu(x)$, which are associated with the links of the lattice,
and the Wilson action is given by 
\be
S \; = \; \beta \sum_{x, \mu > \nu} \, 1 \;-\; \frac{1}{2} \, \tr \; 
P_{\mu \nu }(x)  , 
\en
where the plaquette is defined as 
\be 
P_{\mu \nu }(x) \; = \; 
U_{\mu}(x) \; U_{\nu}(x+e_{\mu}) \; U^{\dagger}_{\mu}(x+e_{\nu}) \; 
U^{\dagger}_{\nu}(x)
\label{eq:plaq}
\en 
and $ e_{\mu} $ is a unit vector in the positive $ \mu $ direction.
(Note that the trace extends over color indices only.)
The Wilson action is invariant under the 
gauge transformation
\be 
U^{\Omega}_\mu(x) \; = \; \Omega(x) \, U_{\mu}(x) \,
 \Omega^{\dagger}(x+e_{\mu}) \; , 
\label{eq:defUtransf} 
\en
where $ \Omega(x) $ are $ SU(2) $ matrices.
The link variables $ U_\mu(x) $ may be expressed in terms of
the continuum
gauge field $A_{\mu}(x)$ by making use of the relation
\be
U_{\mu}(x) \; = \; \exp{\left[ i \, a\, g_0 \, 
                        A^b_{\mu}(x)\, t^b \right]} \; ,
\label{eq:UandA}
\en
where $a$ is the lattice spacing, $ t^{b} = \sigma^{b} /2$ are
the generators of the SU(2) algebra and
$\sigma^{b}$ are the Pauli matrices.
One can check that in the naive continuum limit $a \rightarrow 0$
the Wilson action reproduces the continuum action in eq.\ (\ref{action})
if 
\be 
\beta \; = \; 4 / g_0^2 \; = \; 1/ \left(\pi \, 
   \alpha_{0}\right) \; ,
\label{eq:beta}
\en
where $\alpha_0$ is the bare coupling constant (squared).

\vskip 0.3cm 
For the gauge group SU(2), the link variables $U_\mu(x)$ can be given 
in terms of (real) four-vectors of unit length 
\be 
U_\mu(x) \; = \; u_\mu^0(x) \1 \, + \, i \, \vec{u}_\mu(x) \, 
\vec{\sigma} \; , \hbo \left[ u^0_\mu(x) \right]^2 \, + \, 
\left[ \vec{u}_\mu(x) \right]^2 \; = \; 1 \; , 
\label{eq:defU} 
\en 
where $\1$ is a $2 \times 2$ identity matrix.
By defining the lattice gluon field ${\cal A}^b_{\mu}(x)$ as
\be
{\cal A}^b_{\mu}(x) \;=\;
  \frac{U_\mu(x) \,-\, U_{\mu}^{\dagger}(x)}{2 \, i}
\label{eq:standard_gluon}
\en
one obtains
\be
{\cal A}^b_{\mu}(x) \;=\; 2 \, u^b_{\mu}(x) \;=\;
    a\, g_0 \, A^b_{\mu}(x) \; + \; {\cal O}(a^3)
\label{eq:standA} 
\; 
\en
in the naive continuum limit $a \to 0$.
This definition has been used for the numerical simulations
done in S\~ao Carlos.

\vskip 0.3cm 
Note that the gluon field ${\cal A}^{b}_{\mu}(x)$ defined above
changes sign under a non-trivial center transformation $ Z_2 $
of the link fields
$ U_{\mu}(x) \rightarrow - \, U_{\mu}(x)$. 
Recently, another identification of the gluonic 
degrees of freedom in the lattice formulation
was proposed \cite{Langfeld:2001cz}. In this case, one
first notices that the gluon field 
in continuum Yang-Mills theories transforms under the {\it adjoint} 
representation of the $SU(2)$ color group, i.e.\
\bea 
A^{a\, \prime}_{\mu}(x) &=& O^{ab}(x) \, A^b_{\mu}(x) 
\, + \, \frac{\epsilon^{aed}}{2} \, O^{ec}(x) \, \partial_{\mu} O^{dc}
\label{eq:Aadjo} \\[0.2cm] 
O^{ab}(x) &=& 2 \, \Tr\, \left[ \,\Omega(x) \, t^a \, 
\Omega^\dagger(x) \, t^b \right] \; , 
\label{eq:defO} 
\ena 
where $\Omega(x) \in SU(2)$ is a gauge transformation of the 
fundamental quark field and $O^{ab}(x) \, \in \, SO(3)$. 
In view of the transformation properties in eq.\ \reff{eq:Aadjo}, one can
identify the continuum
gauge fields $ A_{\mu}^a(x) $ with the algebra-valued fields
of the adjoint representation
\be 
{\cal U}_{\mu}^{cd}(x) \; = \; \left\{\exp{\left[ 
a \, g_0 \, A^{b}_\mu(x) \, \hat{t}^b \right]} \, \right\}^{cd} \; ,
\label{eq:Uadjo} 
\en 
where $ \hat{t}^b_{ac} = \epsilon^{abc} $ and
the total anti-symmetric tensor $\epsilon^{abc}$ 
is the generator of the $SU(2)$ group in the adjoint representation. 
On the lattice, the adjoint links $ {\cal U}_{\mu }^{ab}(x) $ are obtained
from 
\be 
{\cal U}_{\mu }^{cd}(x) \; = \;  2 \, \Tr \left[\, U_{\mu}(x) \, t^c \, 
U^{\dagger}_{\mu}(x) \, t^d \,\right] 
\label{eq:Uadjo2} 
\en
and the gluon field $ {\cal A}_{\mu}^a(x)$ is given by
\be
{\cal A}^b_{\mu}(x)  \; = \;   2 \, u^0_\mu(x)
\, u^b_\mu(x) \; ,
\label{eq:Acoset}
\en
without summation over $\mu$ on the right-hand side.
By expanding eq.\ \reff{eq:Uadjo} in powers of the
lattice spacing $ a $ and by using eqs.\ \reff{eq:defU}
and \reff{eq:Uadjo2} one obtains
\be 
{\cal A}^b_{\mu}(x)  \; = \;   a\, g_0 \, A^b_{\mu}(x) 
\; + \; {\cal O}(a^3) \; .
\en 
Clearly, the representation \reff{eq:Acoset} is invariant
under a non-trivial center transformation $ U_\mu (x) 
\rightarrow - \, U_\mu (x) $. This discretization of the gluon
field has been used for the simulations done in T\"ubingen.

\vskip 0.3cm 
It is well known that different discretizations of the
gluon field lead to gluon propagators
equivalent up to a trivial (multiplicative) renormalization
\cite{Giusti:1998ur}--\cite{Bogolubsky:2002ui}.
Also, this proportionality constant between different discretizations
of the gluon propagator may be (partially) explained as a tadpole
renormalization \cite{Cucchieri:1998ta,Cucchieri:1999ky}.
We point out, however, 
that it is useful to disentangle the information carried by
center elements and coset fields, defined above, when the vacuum 
energy is investigated. In particular, it was found that --- in the
continuum limit --- the center elements provide a contribution to the
gluon condensate \cite{Langfeld:2001fc,Langfeld:2001tj}. 


\subsection{Tadpole improved gluon fields} 

The relation between lattice and continuum gluon fields relies 
on the expansion [see eq.(\ref{eq:UandA})] 
\be 
U_\mu(x)  \; = \; 1 \; + \; i \, a \, g_0 \, A^b_\mu (x) \,
t^b \; + \; \ldots 
\en 
where the ellipses denote higher order terms in the bare
coupling constant $g_0$.
The artificial contributions of these higher order terms to 
loop integrals
are called ``tadpole'' terms \cite{Lepage:1992xa}. 
These terms are only suppressed by powers of $g_0^2$ and are 
generically large in simulations using moderate $\beta $ values. 

In order to remove the tadpole contributions from the observable 
of interest one can redefine the relation between 
the link matrices and the continuum gluon field by using
\be 
U_\mu(x)  \; = \; u_{0,L} \Bigl[ 
1 \; + \; i \, a \, g_0 \, A^b_\mu (x) \,
t^b \; + \; \ldots \Bigr] \; , 
\label{eq:tad}
\en 
where $u_{0,L}$ is given by the ``meanfield'' value
\be
u_{0,L} \; = \; \Bigl\langle \frac{\trc}{2} \, U_\mu(x)  \Bigr\rangle \;  
\hbox to 6cm {\hfil (for arbitrary $\mu $)  } 
\en
with the links $ U_\mu(x) $ fixed to the Landau gauge.
Equivalently, one can use \cite{Lepage:1992xa} a gauge invariant 
definition of the tadpole factor given by
\be
u_{0,P} \; = \; \biggl[ \Bigl\langle  \frac{\trc}{2} \, P_{\mu \nu}(x)  
\Bigr\rangle \, \biggr]^{1/4} \; , 
\label{eq:u0stand}
\en
where $ P_{\mu \nu }(x) $ is the plaquette defined in eq.\
(\ref{eq:plaq}). 
Thus, the use of tadpole improvement in
the case of the standard definition of the lattice gluon
field \reff{eq:standA} gives
\be
{\cal A}^b_{\mu}(x)  \; = \;  a\, g_0 \, A^b_{\mu}(x) \; + \; {\cal O}(a^3) 
\; = \;  2 \, u^b_\mu(x) \; / \; u_{0,P} \; .
\label{eq:tadf}
\en

\vskip 0.3cm 
In the case of the coset definition (\ref{eq:Acoset}) of the
gauge fields, the tadpole factors are expressed in terms of
the expectation values of the adjoint link and of the 
adjoint plaquette and are given by
\bea 
u^{\mathrm{ad}}_{0,L} &=& \Bigl \langle \frac{ \trc }{3} \, 
{\cal U}_{\mu }(x)  
\Bigr\rangle \; , 
\\
u^{\mathrm{ad}}_{0,P} &=&  \biggl[ \frac{1}{3}  \Bigl\langle \trc \, 
{\cal P}_{\mu \nu } (x)  \Bigr\rangle \biggr]^{1/4} 
\; = \;  \biggl[ \frac{1}{3} \Bigl\langle
[ \trc  P_{\mu \nu }(x) ]^2 \, - \, 1 \Bigr\rangle \biggr]^{1/4} . 
\label{eq:tadfa}
\ena
Thus, the tadpole improved relation between the continuum gauge field 
$A_\mu ^b (x)$ and the link matrices is given in this case by 
\be
{\cal A}^b_{\mu}(x)  \; = \;  a\, g_0 \, A^b_{\mu}(x) \; + \; {\cal O}(a^3) 
\; = \; 2 \, u^0_\mu(x) \, u^b_\mu(x) \; / \; u^{\mathrm{ad}}_{0,P} 
\; . 
\label{eq:tada}
\en

\vskip 0.3cm
In the sections below we will stress the effect of tadpole improvement
on the various quantities considered in this work.


\subsection{Minimal Landau Gauge}
\label{gaugefix}

The gluon and ghost propagators depend on the choice of the gauge.
In order to maintain contact with the Dyson-Schwinger approach
and the results presented
in the Introduction, we consider the so-called minimal (lattice)
Landau gauge. This gauge condition is imposed by minimizing the
functional
\be
S_\mathrm{fix}[\Omega] \; = \; -
\sum_{x, \mu} \Tr \, U^{\Omega}_\mu(x)
 \; ,
\label{eq:minfunct}
\en
where $U^{\Omega}_{\mu}(x)$ is the gauge-transformed link (\ref{eq:defUtransf}).
This minimizing condition corresponds to imposing the transversality
condition
\be
\left(\Delta \cdot {\cal A}\right)^b(x) \;=\; \sum_{\mu}\,
   {\cal A}^{b}_{\mu}(x)\,-\,{\cal A}^{b}_{\mu}(x - e_{\mu}) = 0
\hbo \forall \; b \, \;\mbox{and}\; x \; ,
\label{eq:divA}
\en
which is the lattice formulation of the usual Landau gauge-fixing condition
in the continuum. Let us notice that the condition \reff{eq:divA}
is exactly satisfied by the lattice gauge field only if the
standard discretization \reff{eq:standA}
is considered, while for the discretization given in \reff{eq:Acoset}
the above result is valid up to discretization errors of order $ a^2 $.
However, in both cases, the gauge-fixing condition \reff{eq:minfunct}
implies that the {\it continuum} Landau gauge-fixing condition
$ \partial \cdot A = 0 $ is satisfied up to discretization errors of order
$ a^2$. In practice, we stop the gauge fixing when the average value of
$[ (\Delta \cdot {\cal A})^b(x) ]^{2}$ is smaller than $10^{- 12}$.

The minimizing condition \reff{eq:minfunct} also implies
that the Faddeev-Popov matrix is positive semi-definite. In particular,
the space of gauge-fixed
configurations $ \{ U^{\Omega}_{\mu}(x) \} $
lies within the first Gribov horizon, where the smallest
(non-trivial) eigenvalue of the Faddeev-Popov operator is zero.
It is well known that, in general, for a given lattice configuration
$ \left\{ U_{\mu}(x) \right\}$, there are many possible gauge
transformations $ \Omega(x) $ that correspond to different local minima
of the functional \reff{eq:minfunct}, i.e.\ there are {\it Gribov copies}
inside the first Gribov horizon \cite{Zwanziger:1990by,Zwanziger:gz}.
Thus, the minimizing condition given in eq.\ \reff{eq:minfunct} is not
sufficient to find a unique representative on each gauge orbit. 
A possible solution to this problem is to
restrict the configuration space of gauge-fixed fields 
$ U^{\Omega}_{\mu}(x) $ to the so-called {\it fundamental modular
region} \cite{Zwanziger:1993dh}, i.e.\ to consider for each
configuration $ \left\{ U_{\mu}(x) \right\} $ the {\em
absolute minimum} of the functional \reff{eq:minfunct}.
From the numerical point of view
this is a highly non-trivial task, corresponding to finding the
ground state of a spin-glass model \cite{Marinari:1991zv}.
On the other hand, if local minima are considered,
one faces the problem that different numerical
gauge-fixing algorithms yield different sets of local minima, i.e.\
they sample different configurations from the region delimited by
the first Gribov horizon (see for example \cite{Giusti:2001xf} and
references therein). This implies that numerical results using
gauge fixing could depend on the gauge-fixing algorithm, making
their interpretation conceptually difficult. 

In the simulations done in S\~ao Carlos, a stochastic
overrelaxation algorithm \cite{Cucchieri:1995pn}--\cite{Cucchieri:2003fb} 
was used.
The simulations done in T\"ubingen have employed 
the simulated annealing technique described in detail in
Ref.~\cite{Suman:1995zg}. 
The problem of Gribov copies was not considered in either case.
Even though neither method is able to locate the {\em global} minimum of 
the gauge-fixing functional, i.e.\ to restrict the gauge-fixed configuration
space to the fundamental modular region, a comparison of the
propagators obtained using the two methods can provide an
estimate of the bias ({\it Gribov noise}) introduced by the gauge-fixing
procedure. From this comparison we have found that the data for the propagators
are rather insensitive to the particular choice of gauge-fixing
algorithm, suggesting that the influence of Gribov copies on the
two propagators (if present) is at most of the order of magnitude of the
numerical accuracy. For the gluon propagator this result is in agreement
with previous studies in Landau gauge for the $SU(2)$
and $SU(3)$ groups in three \cite{Cucchieri:2003di} and four dimensions
\cite{Cucchieri:1997dx,Giusti:2001xf,Mandula:nj}.
A similar result has also been obtained for the gluon propagator in
Coulomb gauge \cite{Cucchieri:2000gu}.
On the other hand, a previous study of the
ghost propagator in $SU(2)$ Landau gauge \cite{Cucchieri:1997dx}
has shown a clear bias related to Gribov copies in the strong-coupling regime.
In particular, data (in the IR region) obtained considering only absolute
minima have been found to be systematically smaller than data obtained using
local minima.
This result --- which has been recently confirmed in \cite{Bakeev:2003rr} 
--- can be qualitatively explained. In fact, as said above, the
smallest non-trivial eigenvalue $\lambda_{min}$ of the Faddeev-Popov operator
goes to zero as the first Gribov horizon is approached. At the same time,
one expects that global minima (i.e.\ configurations belonging to the
fundamental modular region)
be ``farther'' away from the first Gribov horizon than local minima. Thus,
the absolute minimum configuration should correspond to a value of
$ \lambda_{min} $ larger --- on average --- than the value obtained in
a generic relative minimum.\footnote{~This was
checked numerically in Ref.\ \protect\cite{Cucchieri:1997ns}.}
Since the ghost propagator is given by the inverse of the Faddeev-Popov matrix
(see Section \ref{ghost} below), this would imply a smaller ghost propagator 
(on average) at the absolute minimum, 
as observed in Refs.\ \cite{Cucchieri:1997dx,Bakeev:2003rr}.
The analysis carried out in these references has
shown that Gribov-copy effects are visible only for the smallest nonzero
momentum on the lattice, at least for the lattice volumes considered, which
are still relatively small. As explained below (see Section \ref{sec:factors}),
in our analysis we have not considered the data points
corresponding to the smallest momenta.


\subsection{Gluon propagator}
\label{sec:gp}

The continuum gluon propagator in position space is given by
\be 
D^{ab}_{\mu \nu}(x-y) \; = \; \bigl\langle A^a_\mu(x) \, 
A^b_\nu(y) \, \bigr\rangle \; . 
\label{eq:Dofxc} 
\en 
Correspondingly, one can consider the (position space)
lattice gluon propagator
\be 
{\cal D}^{ab}_{\mu \nu}(x-y)
   \; = \; \bigl\langle {\cal A}^a_\mu(x) \, 
{\cal A}^b_\nu(y) \, \bigr\rangle \; , 
\label{eq:Dofx} 
\en 
where ${\cal A}^a_\mu(x)$ is one of the lattice discretizations 
of the continuum gluon field discussed above
[see eqs.\ (\ref{eq:standA}) and (\ref{eq:Acoset})]. 
At the leading order $a$ these two quantities are related by 
\be 
g_0^2 \, a^2 \, D^{ab}_{\mu \nu}(x-y) \; = \; {\cal D}^{ab}_{\mu
  \nu}(x-y) \, / \, u^2_{0,P} \; ,
\label{eq:rela} 
\en 
where $u_{0,P}$ is the tadpole factor given in eq.\
\reff{eq:u0stand} [respectively eq.\ \reff{eq:tadfa}] when considering
the lattice gluon field defined in eq.\
\reff{eq:standA} [respectively eq.\ \reff{eq:Acoset}].

The lattice gluon propagator in momentum space is obtained by
evaluating the Fourier transform
\be 
 {\cal D}_{\mu \nu }^{ab}(\hat{p}) \; = \; \frac{1}{V} \sum_{x, y} 
 {\cal D}_{\mu \nu}^{ab}(x-y) 
\; \exp{\left[ i\, \hat{p}\cdot (\hat{x}- \hat{y}) \right]} \; , \hbo 
\hat{p}_{\mu} = \frac{2 \,\pi\, n_{\mu}}{N_{\mu}} \; , 
\label{eq:Dofp} 
\en 
where $n_{\mu}$ labels the Matsubara modes in the $ \mu $ direction,
$N_{\mu}$ is the number of lattice points in the same direction,
$x = \hat{x} a$, $ y = \hat{y} a$ and 
$V$ is the lattice volume. 
In order to minimize discretization effects \cite{Marenzoni:1994ap},
we consider the gluon
propagator as a function of the lattice momentum $p$ with components
\be 
p_{\mu} \; = \; 2 
\, \sin{\left(\frac{\hat{p}_{\mu}}{2} \right)} \; .
\label{eq:defplatt} 
\en 
It is also useful to introduce the lattice gluon form factor ${\cal F}
(\hat{p}^2)$, defined
as
\be 
 {\cal D}(\hat{p}) \; = \; \frac{{\cal F}(\hat{p}^2)}{p^2} \; , \hbo 
 {\cal D}(\hat{p}) \; = \; \frac{1}{9} \sum_{a, \mu}  
 {\cal D}^{aa}_{\mu \mu}(\hat{p}) \; ,
\label{eq:Fdef} 
\en 
which is a measure of the deviation of the full
propagator from the free one. Note that, in Landau gauge, the propagator
is diagonal in color space and transversal in Lorentz space. The 
transversality condition \reff{eq:divA} implies \cite{Cucchieri:1997dx}
that ${\cal D}(0)$
is not given by ${\cal D}(\hat{p})$ at $\hat{p} = 0$.
In fact, for $\hat{p} = 0$
the previous equation becomes ${\cal D}(0) = 1/12 \sum_{a, \mu}
{\cal D}^{aa}_{\mu \mu}(0)$.

In order to evaluate numerically the gluon propagator in
momentum space it is useful to employ the formula
\cite{Cucchieri:1997dx}
\bea
 {\cal D}(\hat{p}) & = & \frac{1}{9 V}\, \sum_{a, \mu} \, \Bigl\langle \,
\left[ \sum_{x}\, {\cal A}^a_\mu(x) \, \cos{(\hat{p}\cdot \hat{x})} 
\, \right] ^2
\nonumber \\[0.2mm]
& & \qquad \quad \;\;\; + \, \left[ \sum_{x}\, {\cal A}^a_\mu(x) \,
\sin{(\hat{p} \cdot \hat{x})} \, \right]^2 \, \Bigr\rangle \; .
\label{eq:Dofpnum}
\ena
In fact, by expanding the previous equation we obtain
\be
 {\cal D}(\hat{p}) \;=\; \frac{1}{9 V}\, \sum_{a, \mu} \sum_{x, y} \,
   \Bigl\langle \; {\cal A}^a_{\mu}(x) \, {\cal A}^a_{\mu}(y)  \;
\Bigr\rangle \;\cos{\left[ \hat{p} \cdot (\hat{x} - \hat{y}) \right]}
\; ,
\label{eq:Dofpnum2} 
\en 
which is directly related to eq.\ \reff{eq:Dofp}.

One can also evaluate the form factor $F(\hat{p}^2)$ directly
\cite{Langfeld:2001cz}. To this end we can consider,
without any loss of generality,
a momentum transfer parallel to the time direction
$\hat{p} = (0,0,0,\hat{p}_4)$ and define
\be 
\Delta_t {\cal A}_{\mu}(x) \; = \; {\cal A}_{\mu}(x + e_4)
\, - \, {\cal A}_{\mu}(x) \; ,  
\label{eq:diffA} 
\en 
where $e_4$ is the unit vector in the time direction. The form
factor is then obtained from 
\bea 
\!\!\!\!\! {\cal F}(\hat{p}^2) & = & \frac{1}{9 V} \, \sum_{a, \mu} \, 
\Bigl\langle \, \left[ \sum_{x} \Delta_t {\cal A}^a_{\mu}(x) 
                      \, \cos{(\hat{p}\cdot \hat{x})} \, \right]^2
\nonumber \\ 
& & \qquad \quad \;\; + \, \left[ \sum_{x} \Delta_t {\cal A}^a_\mu(x) \, 
\sin{(\hat{p}\cdot \hat{x})} \, \right]^2 \; \Bigr\rangle \; . 
\label{eq:Fofpnum} 
\ena 
By expanding the previous formula one can verify that
the free part $ 1/p^2 $ is canceled exactly. This
strongly suppresses the statistical noise in the high-momentum
regime. Here, we will present results that directly address the 
gluon propagator \reff{eq:Dofpnum} and the gluon form factor 
\reff{eq:Fofpnum}, evaluated respectively in S\~ao Carlos and in
T\"ubingen.


\subsection{Ghost propagator}
\label{ghost}

The ghost propagator $ G^{ab}(\hat{p}) $ is uniquely defined once the
gauge-fixing functional \reff{eq:minfunct} is specified. In fact,
if we write the gauge-fixing matrix as
\be 
\Omega(x) \; = \; \exp \left[ i \,\theta^a(x) \, t^a \right] \; , 
\label{eq:omegadec} 
\en 
with $ t^a $ defined as in Section \ref{sec:gluonfield},
then the gauge-fixing functional can be expanded with respect to 
the angles $ \theta^a(x) $ and, at any local minimum of $ S_\mathrm{fix} $,
we obtain
\be 
S_\mathrm{fix} \; = \; S_0 \; + \; \frac{1}{2} \sum_{x, y} \sum_{a, b} \,
\theta^a(x) \; M^{ab}_{xy} \; \theta ^b (y) \; + \; {\cal O}(\theta^3) \; , 
\en 
where $ M^{ab}_{xy} $ is the so-called Faddeev-Popov operator. 
Note that the linear term in $ \theta(x) $ is absent by virtue of the minimizing
gauge-fixing condition [see eqs.\ \reff{eq:minfunct} and
\reff{eq:divA}]. The expression of the Faddeev-Popov
operator in terms of the gauged-fixed link variables can be found in
\cite[eq.\ (B.18)]{Zwanziger:1993dh}.
Note that the matrix $M^{ab}_{xy}$
obtained in this way is a lattice discretization of the continuum
Faddeev-Popov operator $ (-\partial + A)\cdot\partial $ and that
this discretization yields automatically the standard discretization 
$ {\cal A}^b_{\mu}(x) $ for the gluon field given in eq.\ \reff{eq:standA}.

The lattice ghost propagator $ {\cal G}^{ab}(\hat{p}) $
is provided by the inverse Faddeev-Popov operator
$M^{ab}_{xy}$. Due to translation invariance, the lattice average 
of the inverse operator depends only on $ (x - y) $. Thus,
in momentum space we have
\be 
{\cal G}^{ab}(\hat{p}) \; = \; \frac{1}{V} \sum_{x, y} \langle 
\left(M^{-1}\right)^{ab}_{xy}  \rangle \; \exp \left[- i\, 
\hat{p} \cdot \left(\hat{x} - \hat{y}\right) \right] \;  . 
\label{eq:Gofp} 
\en 
Since the matrix $M^{ab}_{xy}$ depends linearly on the link
variables $U_{\mu}(x)$, tadpole improvement applied to the ghost
propagator implies a rescaling
\be
{\cal G}^{ab}(p) \;\to\; {\cal G}^{ab}(\hat{p}) \; u_{0,P}
\; .
\en
Thus, at the leading order $a$ one has
\be
a^2 \, G^{ab}(p)\;=\; {\cal G}^{ab}(\hat{p}) \; u_{0,P}
\; ,
\label{eq:relaG}
\en 
where $G^{ab}(p)$ is the continuum ghost propagator in
momentum space.

The asymptotic behavior of the ghost propagator $ G^{ab}(\hat{p}) $ is known
from perturbation theory: it decreases as $ 1/p^2 $ with additional 
logarithmic corrections. The $ 1/p^2 $ behavior is inherited from 
the free-theory case. The non-trivial information on the ghost
propagator is therefore encoded in the (continuum)
form factor $ J(\hat{p}^2) $, which is defined by 
\be 
G^{ab}(\hat{p}) \; = \; \delta ^{ab} \; G(\hat{p}) \; = \; \delta ^{ab}
 \frac{J(\hat{p}^2)}{p^2} \; , 
\en
yielding the lattice form
\be
{\cal G}^{ab}(\hat{p}) \; = \; \delta ^{ab} \; 
 \frac{{\cal J}(\hat{p}^2)}{p^2} \; .
\label{eq:defG} 
\en 
Numerically, the lattice ghost propagator can be obtained by
inverting the Faddeev-Popov matrix $M^{ab}_{xy}$. In the numerical simulations
carried out in S\~ao Carlos this has been done using a conjugate-gradient
algorithm. On the contrary, in the simulations in T\"ubingen the
(lattice) ghost form factor ${\cal J}(\hat{p}^2)$ has been evaluated directly.
To this end one can consider
the following set of linear equations (for a given set of link variables $U$)
\bea
\sum_{y, b} \; M^{ab}_{xy}[U] \; \bar{u}^b(y) &=& n^a \, 
\left\{ \cos{\left[\hat{p}\cdot (\hat{x}-e_{\mu})\right]} \, - \, 
        \cos{\left[\hat{p}\cdot \hat{x}\right]} \right\}  
\label{eq:eqM1} \\ 
\sum_{y, b} \; M^{ab}_{xy}[U] \; \bar{v}^b(y) &=& n^a \, 
\left\{ \sin{\left[\hat{p}\cdot (\hat{x}-e_{\mu})\right]} \, - \, 
        \sin{\left[\hat{p}\cdot \hat{x}\right]} \right\} \; , 
\label{eq:eqM2} 
\ena 
where $n^a$ is an arbitrary unit vector which specifies
the components of the ghost propagator under investigation. 
We are considering momenta with a non-zero
component only in the $ \mu $ direction.
In fact, by solving these equations for $ \bar{u}^b(y) $ and
$ \bar{v}^b(y)$ and using trigonometric identities we find
that the ghost form factor is given by 
\bea 
\!\!\!\!\! {\cal J}(\hat{p}^2) & = & \frac{1}{V} \, \sum_y \, \biggl\langle 
 \left\{ \cos{\left[\hat{p}\cdot (\hat{y}-e_{\mu})\right]} \, - \, 
        \cos{\left[\hat{p}\cdot \hat{y}\right]} \right\} \, n^b 
\, \bar{u}^b(y) 
\label{eq:calJ1} \\
& & \qquad \quad + \, \left\{ \sin{\left[\hat{p}\cdot 
(\hat{y}-e_{\mu})\right]} \, - \,\sin{\left[\hat{p}\cdot
    \hat{y}\right]} 
\right\} \, n^b \, \bar{v}^b(y) \, \biggr\rangle 
\nonumber \\[0.2cm] 
&=&  \frac{1}{V} \,\sum_{x, y} \sum_{a, b} \, n^b \,
  \langle \left(M^{-1}\right)^{ab}_{xy} \rangle \, n^a \, 
    \left[4 \, \sin^2\left(\frac{\hat{p}_{\mu}}{2}\right) \right] \,
\cos{\left[\hat{p}\cdot (\hat{x} - \hat{y})\right]} \; .
\label{eq:calJ2} 
\ena 
Note that, with our choice of momenta, the lattice momentum squared is given by
$ p^2 = 4 \sin^2(\hat{p}_{\mu}/2) $, implying that the free part $ 1/p^2 $ of
the ghost propagator exactly cancels out and we are left with the form factor
${\cal J}(\hat{p}^2)$. 
The set of equations \reff{eq:eqM1}--\reff{eq:eqM2} has been solved using
a bi-conjugate gradient method for matrix inversion. 


\subsection{Renormalization} 
\label{renorm}

Renormalization of Yang-Mills theories in four dimensions implies
that the bare coupling acquires a dependence on the
ultraviolet (UV) cutoff $\Lambda_{UV}$ given by
\be 
\alpha_0 \rightarrow \alpha_0( \Lambda_{UV}/ \Lambda_{scale}) \, . 
\en 
Thereby, the bare coupling constant is no longer the theory's parameter.
The Yang-Mills scale parameter $ \Lambda_{scale} $ 
takes over the role of the only parameter of the theory. 
In the context of (quenched) lattice gauge simulations the string 
tension $\sigma $ is widely used as the generic low-energy scale. 
In this case, the cutoff dependence of the bare 
coupling is implicitly given 
by the $\beta $ dependence of $\sigma \, a^2( \beta ) $ 
where $\beta $ is related to the bare coupling in (\ref{eq:beta}) and 
$\Lambda _{UV} = \pi/a(\beta )$. 
In Section \ref{sec:mom} we will derive this relation from lattice data
(obtained in Ref.\ \cite{Fingberg:1992ju}).

\vskip 0.3cm
In addition, wave-function renormalization constants develop a
dependence on $\Lambda_{UV}/ \Lambda_{scale}$. 
The lattice bare form factors of the previous subsections,
${\cal F}_B$ and ${\cal J}_B$, 
are related to their continuum analogues (for very large $\beta$) by 
\be
F_B \; = \; {\cal F}_B \; \frac{\beta }{ 4 \, u_{0,P}^2 } \; , \hbo 
J_B \; = \; {\cal J}_B \; u_{0,P} \; . 
\en 
These form factors depend on the momentum $p^2$ 
and on the UV cutoff $\Lambda_{UV}$ (given in units of the
string tension) or, equivalently,
on the lattice coupling $\beta$ (see Section \ref{sec:factors}
below). Thus, we can write 
\be
{\cal F}_B \; = \; {\cal F}_B(p^2, \beta) \; , \hbo 
{\cal J}_B \; = \; {\cal J}_B(p^2, \beta) \; . 
\en 
The renormalized form factors are obtained upon multiplicative 
renormalization
\bea 
F_R(p^2, \mu^2) &=& Z_3^{-1}(\beta, \mu) \; F_B(p^2, \beta) 
\label{eq:FR} \\
J_R(p^2, \mu^2) &=& \widetilde{Z}_3^{-1}(\beta, \mu) \; J_B(p^2, \beta)
\label{eq:JR}
\ena 
using the renormalization conditions 
\be 
 F_R (\mu^2, \mu ^2)  \; = \; 1 \; , \hbo 
 J_R (\mu^2, \mu ^2)  \; = \; 1 \; . 
\label{eq:renorcond} 
\en 
(Notice that tadpole renormalization does not affect the calculation
of $F_R$, $J_R$ but may be useful for the determination of the renormalization
constants $Z_3$, $\widetilde{Z}_3$.)

Clearly, similar relations hold also for the bare and renormalized gluon and
ghost propagators.
In practice, the multiplicative renormalizability of the theory 
implies that a rescaling of the data for each $ \beta $ value (independently
of the lattice momentum) is sufficient to let 
the form factors $ {\cal F}_B(p^2, \beta) $ and $ {\cal J}_B(p^2, \beta) $
--- or equivalently the corresponding propagators --- fall on top 
of a single curve describing the momentum dependence of the 
corresponding renormalized quantity.


\subsection{Running coupling constant}
\label{sec:runcoup}

Of great importance for phenomenological purposes is the running coupling 
strength $\alpha_R(p^2)$ 
considered in Ref.~\cite{vonSmekal:1997is}. In particular,
this strength enters directly the quark DSE and can be interpreted
as an effective interaction strength between quarks \cite{Bloch:2002eq}.
This running coupling
strength is a renormalization-group-invariant combination of the gluon and
ghost form factors. In order to derive this combination we can start with 
the definition of the ghost-ghost-gluon-vertex renormalized coupling strength 
\be 
\alpha_R(\mu^2) \; = \; \frac{Z_3(\beta, \mu)  \;
\widetilde{Z}^2_3(\beta, \mu)}{\widetilde{Z}^2_1(\beta, \mu)}
\; \alpha_{0} (\Lambda _{UV}) \; , 
\label{eq:alphaR} 
\en 
where $\widetilde{Z}_1(\beta, \mu)$
is the ghost-ghost-gluon-vertex renormalization constant.
In lattice simulations, the UV-cutoff is related to $\beta $ 
by $\Lambda _{UV} = \pi / a(\beta )$, where $a$ is the lattice
spacing. Using eqs.\ \reff{eq:FR} and
\reff{eq:JR} we can express the renormalization constants 
$ Z_3(\beta, \mu) $ and $ \widetilde{Z}_3(\beta, \mu)$ in terms 
of the bare and renormalized form factors yielding 
\be 
\alpha_R(\mu^2) \; F_R(p^2, \mu^2) \; J^2_R(p^2, \mu^2) \; = \; 
\frac{\alpha_{0} (\Lambda _{UV}) }{\widetilde{Z}^2_1(\beta, \mu)} \; 
F_B(p^2, \beta) \; J^2_B(p^2, \beta) \; . 
\label{eq:alphaRandB} 
\en 
Note that the left-hand side of this relation is finite and 
independent of $\beta $ 
by construction and that the
right-hand side depends on the renormalization scale $\mu$
only through the
ghost-ghost-gluon-vertex renormalization constant
$\widetilde{Z}_1(\beta, \mu)$.
It was found more than twenty years ago by Taylor \cite{Taylor:ff}
that (in the continuum) $ \widetilde{Z}_1$ is finite, 
independent of $\mu $, at least to all orders of perturbation theory. 
This finding will be confirmed by our lattice studies below. 
In this case, the right-hand side of eq.\ \reff{eq:alphaRandB}
is thus independent of $\mu$.
Then, by choosing $\mu = \sqrt{p^2}$ and using the
renormalization conditions \reff{eq:renorcond},
we find the final expression 
for the running coupling strength, i.e., 
\be 
\alpha_R(p^2) \; = \; 
\alpha_R(\mu^2) \; F_R(p^2,\mu ^2) \; J_R^2(p^2, \mu^2) \; . 
\label{eq:alphaRfin} 
\en 

Finally, let us notice that eqs.\ \reff{eq:rela} and \reff{eq:relaG}
imply that tadpole renormalization does not affect the renormalized
coupling defined above.


\section{Details of the numerical simulations} 
\label{sec:details}


\subsection{Setup}
\label{sec:setup}

All our simulations used the standard Wilson action for
$SU(2)$ lattice gauge theory in four dimensions with
periodic boundary conditions. In order to check
finite-volume effects and verify scaling we consider
several values of $\beta$ and of the
lattice volumes $V = N_s^3 \times N_t$. The 
dependence of the lattice spacing on $\beta$ can be inferred
from a calculation of the string tension $\sigma$ in lattice
units. Here, we will use the data for $\sigma a^2$ 
reported in Ref.~\cite{Fingberg:1992ju} and a linear interpolation
of the logarithm of these data where needed.\footnote{~For 
$\beta = 2.15$ an extrapolation of these data was necessary 
in order to obtain $\sigma a^2$.} 
A value of the lattice spacing in physical units was
obtained using the value $\sigma = [440 \, \mathrm{MeV}]^2$
for the string tension.
We used $N_\mathrm{conf}$ independent configurations 
for the numerical evaluation of the propagators.

\begin{table}[t]
\begin{center}
\begin{tabular}{lccccccc} \hline
$\beta $      &       2.2 &      2.3 &       2.4 &
                      2.5 &      2.6 &       2.7 & 2.8 \\ \hline 
$\sigma a^2$  &  0.220(9) & 0.136(2) &  0.071(1) &
                 0.0363(3)  & 0.018(1) &  0.0103(2) & 0.0055(3) \\ \hline 
\end{tabular}
\end{center}
\vskip -5mm
\caption{Simulation parameters of the runs in S\~ao Carlos. 
  Data from Ref.~\cite{Fingberg:1992ju} were used to obtain the lattice 
  spacing in units of the string tension.
  Error bars (in parentheses) come from propagation of errors
  and indicate one standard deviation on the last
  significant digit.
  The lattice volumes $V$ and the
  number of configurations $N_\mathrm{conf}$ considered are discussed
  in the text.
}
\label{tab:1}
\vskip 5mm
\end{table} 
\begin{table}[ht]
\begin{center}
\begin{tabular}{lcccccc} \hline
$\beta $ & $2.15$ & $2.2$ & $2.3$ & $2.375 $ & $2.45$ & 
$2.525$  \\ \hline 
$\sigma a^2$ &  $0.280(13)$ & $0.220(9) $ & $0.136(2)$ & $0.083(2)$ 
& $0.0507(8)$  & $0.0307(5)$ \\ \hline 
$N_s^3 \times N_t$ & $16^3 \times 32$ & $16^3 \times 32$ & 
$16^3 \times 32$ & $16^3 \times 32$ & $16^3 \times 32$ & 
$16^3 \times 32$ \\ \hline 
$N_\mathrm{conf}$ & $200$ & $200$ & $200$ & $200$ & $200$ & 
$200$ \\ \hline 
\end{tabular}
\end{center}
\vskip -5mm
\caption{Simulation parameters of the runs in T\"ubingen. 
  Data from Ref.~\cite{Fingberg:1992ju} were used to obtain the lattice 
  spacing in units of the string tension. }
\label{tab:2}
\vskip 5mm
\end{table} 

Computations in S\~ao Carlos were performed on the PC
cluster at the IFSC-USP (the system has 16 nodes with
866 MHz Pentium III CPU and 256 MB RAM memory). 
All runs in S\~ao Carlos started with a random gauge configuration and for
thermalization we use a {\em hybrid overrelaxed} (HOR) algorithm.
The total computer time used for the runs was about 50
days on the full PC cluster.
In Table \ref{tab:1} we report the parameters used for the simulations 
in S\~ao Carlos. For each $\beta$ value three different
lattice volumes were considered, i.e.\ $V = 14^4, 20^4$ and
$26^4$. For the lattice volume $V = 14^4$ (respectively
$V = 20^4$ and $26^4$) and for each
$\beta$ value we produced $N_\mathrm{conf} = 500$ (respectively
$150$ and $50$) configurations. 

Computations in T\"ubingen were carried out at the local PC
cluster where $4-12$ nodes (1 Ghz Athlon) were used. 
The simulation parameters
for the runs in T\"ubingen are listed in Table~\ref{tab:2}. 

\vskip 0.3cm 
Table \ref{tab:4} lists the differences between the  S\~ao Carlos 
and the T\"ubingen approach. 
Reconstructing the continuum gauge field from the link fields 
in different manners [compare eq.\ (\ref{eq:standA}) with eq.\
(\ref{eq:Acoset})] provide insight into the discretization errors. 
Employing different gauge-fixing algorithms points out the 
effect of the Gribov ambiguities on the propagators. 
 
\vskip 0.3cm 
\begin{table}[h]
\begin{center}
\begin{tabular}{|l|c|c|} \hline
                           &  S\~ao Carlos
                           &  T\"ubingen \cr \hline 
definition of gauge fields & fundamental representation
                           & adjoint representation \cr \hline 
gauge fixing               & iterative stoch.\ overrelaxation
                           & simulated annealing \cr \hline 
number of lattice points   & finite-size control
                           & fixed \cr \hline 
\end{tabular}
\end{center}
\vskip -5mm
\caption{Differences between the  S\~ao Carlos and the T\"ubingen
numerical approach.}
\label{tab:4}
\vskip 5mm
\end{table} 
%


\subsection{Determination of renormalization constants} 
\label{sec:constants}

In order to obtain the renormalized propagators and
form factors, one needs to
evaluate the renormalization constants $ Z_3^{-1}(\beta, \mu) $ and 
$ \widetilde{Z}_3^{-1}(\beta, \mu) $ defined in eqs.\ (\ref{eq:FR}) 
and (\ref{eq:JR}), respectively.
Multiplicative renormalizability implies that 
one can ``collapse'' data obtained at different $\beta$ on a single 
curve. This can be done by using the matching technique described 
in detail in Ref.~\cite[Sec.\ V.B.2]{Leinweber:1998uu}, 
For instance for the gluon form factor this is equivalent to 
considering the quantity
\be
F_R(p^2,\mu^2 ) \;=\; Z_3^{-1} (\beta, \mu ^2) \, F_B(\beta, p^2)
\; ,
\label{eq:fmatch}
\en
where the factor $\, Z_3^{-1}(\beta, \mu)\,$ for each $\beta$ is obtained
from the matching technique. The renormalization point, i.e. the $\mu$ 
dependence, comes into play when the ``single'' curve is rescaled to satisfy 
the condition 
\be 
F_R(\mu^2,\mu^2) \; = \; 1 \; . 
\en 
The same procedure is applied to the ghost form factor. 
For our analysis we considered a renormalization scale
of $\mu = 3 \, $GeV. 

For the S\~ao Carlos data we checked for finite-size effects 
before applying the matching technique.\footnote{~For more
details see \cite[Sec.\ III]{Cucchieri:2003di}.}
In particular, by comparing data at different 
lattice sizes and same $\beta$ value, we find (for each $\beta$) 
a range of momenta for which the data are free from finite-volume corrections.
We then perform the matching using data for these momenta and $V = 26^4$.


\subsection{ Asymptotic scaling } 
\label{sec:mom}

As said in Section \ref{renorm}, the cutoff dependence of the 
bare coupling is implicitly given by $\sigma \, a^2(\beta )$. 
Thereby, the string tension $\sigma $ serves as the fundamental 
energy scale. 
\begin{figure}[t]
\centerline{
\epsfxsize=9.5cm
\epsffile{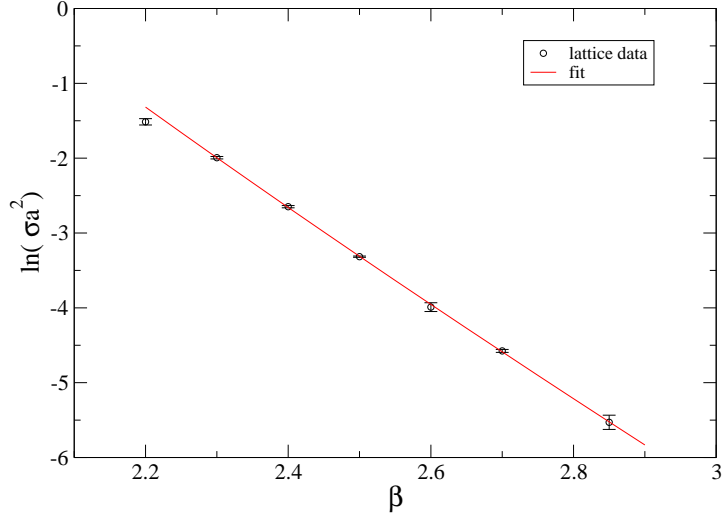}
}
\caption{The string tension in units of the lattice spacing: 
   lattice data from \cite{Fingberg:1992ju} 
   and the fit using eq.\ (\ref{eq:tension}). } 
\label{fig:tension}
\end{figure}
\vskip 0.3cm 
In principle, perturbation theory predicts the $\beta $ 
dependence of $ \sigma a^2$ for the regime $a \ll 1/\sqrt{\sigma}$. 
In practice, large deviations of the measured 
function $\sigma a^2(\beta )$ from the perturbative scaling 
are observed in the scaling region, which corresponds to the interval 
$\beta \in [2.15, 2.8]$ in our case. Clearly, the 
relation between the ``measured'' values for $\sigma a^2$ 
and perturbation theory is highly important for a careful 
extrapolation of the lattice data to continuum physics. 
One goal of the present paper is to present this relation. 

\vskip 0.3cm 
For this purpose, we perform a large $\beta $ expansion of
the lattice spacing in units of the string tension, i.e., 
\be 
\ln \left( \sigma a^2 \right) \; = \; - \; 
\frac{4 \pi^2}{\beta_0} \, \beta \; + \; 
\frac{2 \beta_1}{\beta_0^2} \, \ln \left( \frac{4 \pi^2}{
\beta_0} \, \beta \right) \; + \; \frac{4 \pi^2}{\beta_0} \, 
\frac{d}{\beta} \; + \; c \; .  
\label{eq:tension}
\en 
The first two terms on the rhs of (\ref{eq:tension}) are in 
accordance with 2-loop perturbation theory. The term $ d / \beta $ 
represents higher-order effects and the term $\,c\,$
is a dimensionless scale factor to the string tension. 
Parameters $c$ and $d$ are determined by fitting the formula 
(\ref{eq:tension}) to the lattice data reported in Ref.\ 
\cite{Fingberg:1992ju}. Using only 
data for $\beta \ge 2.3$ we obtain
\be 
c \; = \; 4.38(9) \; , \hbo  d \; = \; 1.66(4) \; , \hbo 
\chi^2 /d.o.f. \; = \; 0.62 \; . 
\label{fitpara} 
\en 
The corresponding fit is shown in Figure \ref{fig:tension}. 
It appears that the truncation of the series (\ref{eq:tension}) at 
the $1/\beta $ level reproduces the measured values to high 
accuracy.

\vskip 0.3cm 
In order to illustrate the impact of the $d$-term correction 
in (\ref{eq:tension}) on the estimate of Yang-Mills scale parameters, 
we briefly consider the lattice scale parameter $\,\Lambda 
_{\mathrm{lat}}\,$. This parameter is implicitly defined at the 2-loop 
level by considering~\cite{Lepage:1992xa} 
\be 
\alpha^{-1}_{\mathrm{lat}} \; = \; \frac{\beta_0}{4 \pi} \, \ln
\left(  
\frac{1}{a^2 \; \Lambda^2_{\mathrm{lat}} }\right)  \; + \; 
 \frac{\beta_1}{2 \pi \beta_0} \, \ln  \left[  
\ln \left( \frac{1}{a^2 \; \Lambda^2_{\mathrm{lat}}} \right) \right] \; , 
\label{eq:lamlat}
\en 
where  for the $SU(N)$ gauge group 
\be
\beta_0 \; = \; \frac{11}{3} \; N \; , \hbo
\beta_1 \; = \; \frac{17}{3} \; N^2  \; ,  \hbo 
\alpha^{-1}_{\mathrm{lat}} \; = \; \frac{2 \pi }{N } \, \beta \; . 
\label{eq:b0b1}
\en 
Inverting eq.\ (\ref{eq:lamlat}) consistently up to 2-loop
perturbation theory yields 
\be 
\ln \left( a^2 \; \Lambda^2_{\mathrm{lat}} \right) 
\; \stackrel{\cdot }{=} \; - \, 
\frac{ 4 \pi }{\beta_0 } \, \alpha^{-1}_{\mathrm{lat}} \; + \; 
\frac{2 \beta _1 }{ \beta_0^2 } \, \ln \left( 
\frac{ 4 \pi }{\beta_0 } \, \alpha^{-1}_{\mathrm{lat}}
\right) \; . 
\label{eq:alam}
\en
Then, using $\alpha^{-1}_{\mathrm{lat}} = \pi \, \beta $ from
eq.\ (\ref{eq:b0b1}) and eliminating $a$ by 
subtracting eq.\ (\ref{eq:tension}) from the latter equation 
we find
\be 
\ln \left( \frac{ \Lambda ^2_{\mathrm{lat}} }{\sigma } \right) \; 
= \lim _{\beta \to \infty } \biggl[ 
\; - \, c \; - \; \frac{ 4 \pi ^2 }{ \beta _0 } 
\frac{d}{\beta } \biggr] \; . 
\label{eq:rat}
\en 
Thus, if we extrapolate to the continuum limit
$\beta \rightarrow \infty $ we obtain
\be 
\Lambda_{\mathrm{lat}} \; = \; e^{- c / 2}\,\sqrt{\sigma}
\;=\; 0.112(5) \, \sqrt{\sigma} \; . 
\label{eq:lamlatnum}
\en 
Using the value $\sigma = [440 \, \mathrm{MeV}]^2$ one gets
$\Lambda_{\mathrm{lat}} \,=\, 49(2)\, \mathrm{MeV}$.
If one instead of the limit in (\ref{eq:rat}) assumes that the 
asymptotic scaling regime is reached for, e.g., $\beta =2.5$, 
one gets $\Lambda_{\mathrm{lat}} = 0.0188(8) \,
\sqrt{\sigma}$. This is the order of magnitude familiar from the
literature. Hence, for a scaling analysis of lattice results 
employing $\beta \in [2.15, 2.8]$ the irrelevant term of order 
$1/\beta$ is still important. 


%
\begin{figure}[t]
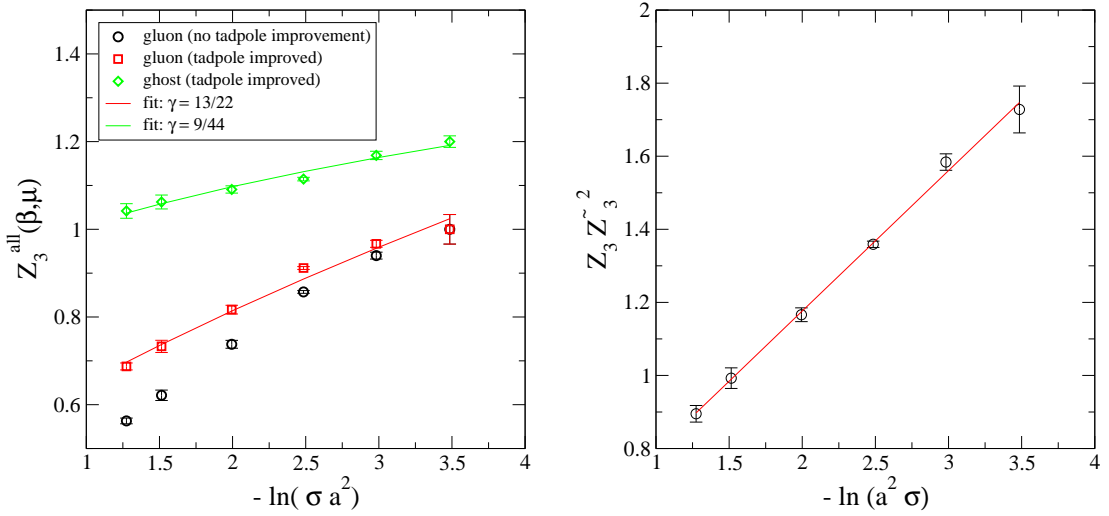

\centerline{
\epsfxsize=7cm
\epsffile{z3_all2.eps} \hspace{.3cm}
\epsfxsize=7cm
\epsffile{z1t.eps}
}
\caption{The gluon and ghost renormalization constants, $Z_3(\beta, \mu)$ and 
 $\widetilde{Z}_3(\beta, \mu)$, for $\mu = 3 \,$GeV (left panel) 
 $y$ axis is arbitrarily scaled. Towards the cutoff dependence of 
 $ \widetilde{Z}_1$  (right panel). Figures corresponding to the
 T\"ubingen data only.}
\label{fig:1}
\end{figure}
\begin{figure}[t]
\vspace{-5.4cm}
\centerline{
\epsfxsize=8cm
\epsffile{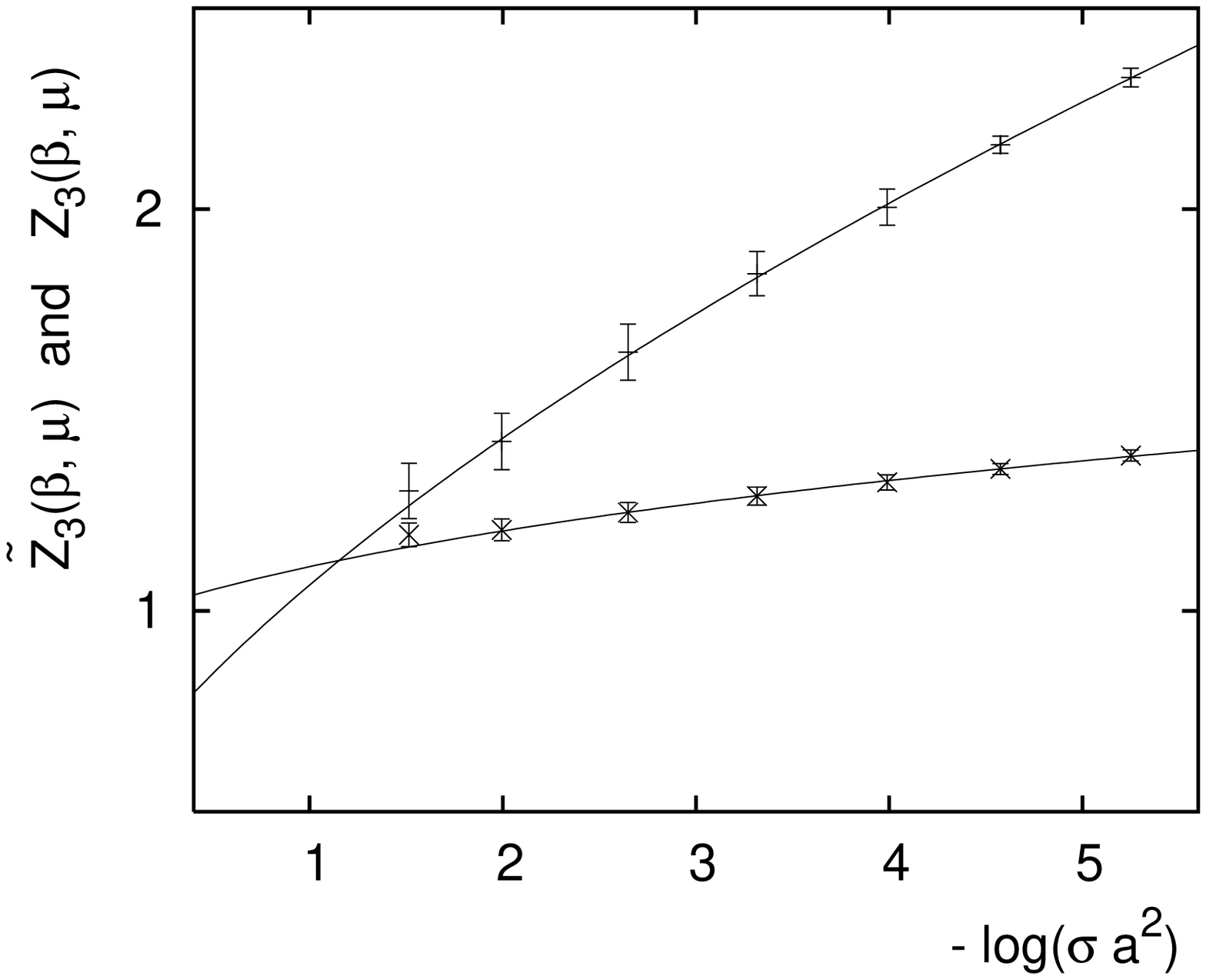}
\epsfxsize=8cm
\epsffile{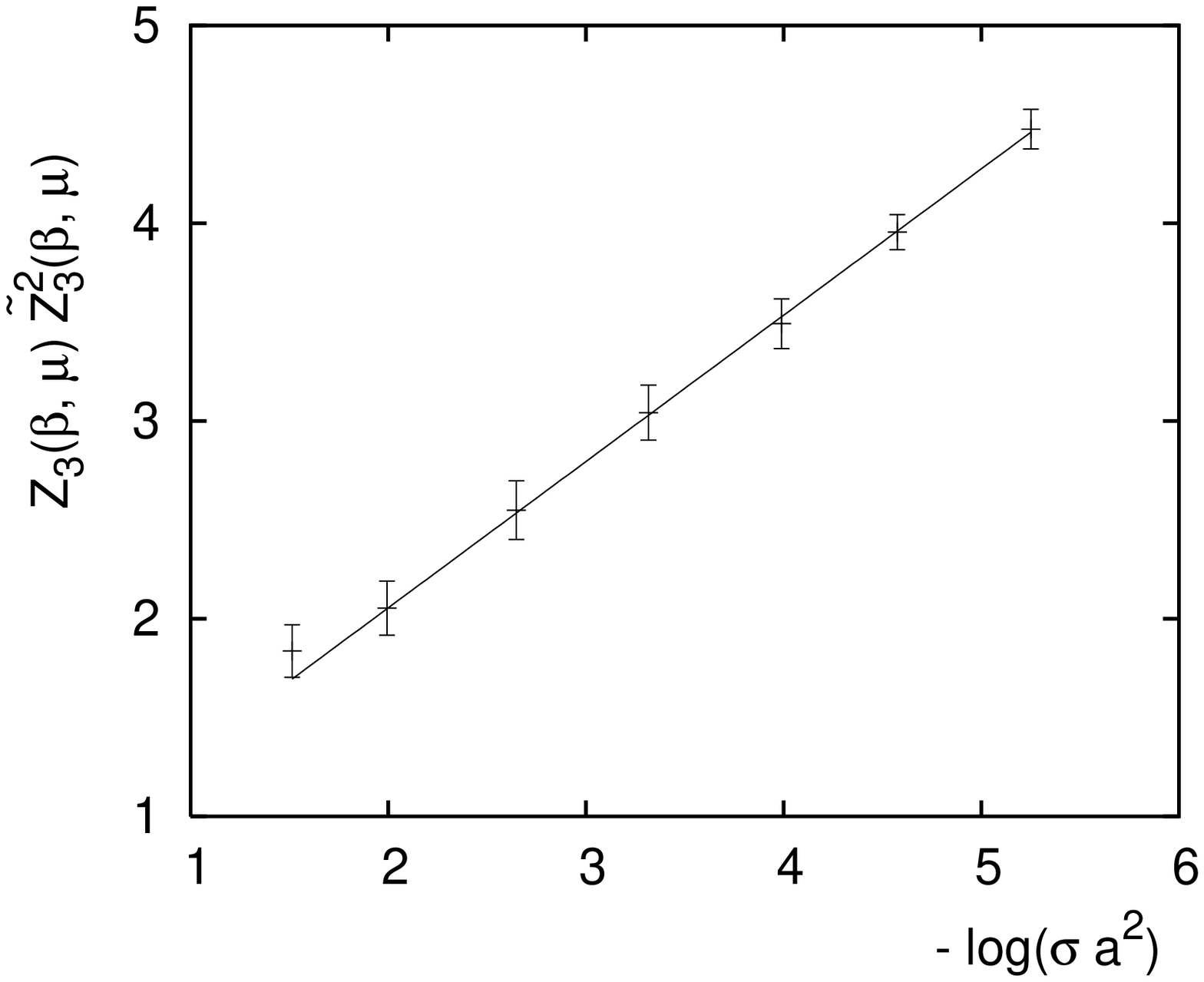}
}
\caption{The analogue of Fig.\ \ref{fig:1}, for the S\~ao Carlos data.
The fits shown on the left-hand side for the gluon (upper curve) and 
ghost (lower curve) renormalization constants are done with $\gamma$ 
as a free parameter. All fits neglect the leftmost data point.}
\label{fig:1B}
\end{figure}
%


\section{Results}
\label{results}


\subsection{Renormalization constants $Z_3$, $\tilde{Z}_3$ and 
$\tilde{Z}_1$}

Let us firstly focus on the renormalization constants
$Z_3$ and $\tilde{Z}_3$. As outlined in Subsection~\ref{sec:constants}, 
these constants are obtained from ``matching'' the lattice data 
from simulations using different $\beta $ values. 
Figures \ref{fig:1} and \ref{fig:1B} show the cutoff dependence of these 
constants respectively for the T\"ubingen and S\~ao Carlos sets of
data.

\vskip 0.3cm 
Using the results above, we can check that our data are consistent with 
the predictions from perturbation theory. For large enough UV cutoff,
one expects that the 1-loop behavior is recovered, i.e.,
\be
Z_3(\beta ,\mu) , \; \widetilde{Z}_3(\beta, \mu)
\; \approx \;   b \; \left[   - \, \ln ( \sigma a^2)  + \omega 
\right]^\gamma \; . 
\label{eq:z3UV}
\en
If only small lattice spacings are considered, $\omega $ is 
related to the ratio between the string tension and the 
Yang-Mills scale parameter at 1-loop level
\be 
\omega \; = \; \ln \frac{ \pi^2 \; \sigma  }{ \Lambda ^2_{1-loop} } \; . 
\en 
Here, we treat $\omega $ as a fit parameter and explore a range 
of lattice spacings where one would already expect significant 
deviations from the 1-loop behavior. 
As shown in Figures \ref{fig:1} and \ref{fig:1B} (left panel), a good 
consistency with the known anomalous dimensions is observed. 

For the T\"ubingen data, we find that 
$\omega \approx 1.13$ is a good choice for reproducing the data for 
$Z_3$ and $ \widetilde{Z}_3$ simultaneously. 
It turns out that within the $\beta $ range explored in the
T\"ubingen runs, tadpole improvement has a minor effect
on the anomalous dimension. The adjoint plaquette used for
the tadpole improvement [see eq.\ (\ref{eq:rela})] employing  the
adjoint representation is listed in Table \ref{tab:3}.
 
\vskip 0.3cm 
\begin{table}[htb]
\begin{center}
\begin{tabular}{lcccccc} \hline
$\beta $ & $2.15$ & $2.2$ & $2.3$ & $2.375 $ & $2.45$ &
$2.525$  \\ \hline
$u^{ad}_{0,P}$ &  $0.225(42)$ & $0.241(43) $ & $0.274(45)$ & $0.297(46)$
& $0.317(46)$  & $0.336(47)$ \\ \hline
\end{tabular}
\end{center}
\vskip -5mm
\caption{ Expectation value of the adjoint plaquette (\ref{eq:tadfa})
  used for the tadpole improvement of the gluon fields derived
  from the adjoint representation. }
\label{tab:3}
\end{table}
 
For the S\~ao Carlos data we have performed the fits with $\gamma$
as a free parameter, leaving out the data point with $\beta=2.2$
(i.e.\ the leftmost point in Figure \ref{fig:1B}). We obtain the 
following values for the gluon and ghost cases
\be
\gamma_{\mathrm{gluon}} \,=\, 0.60(5) \quad \mbox{and}\;\quad
\gamma_{\mathrm{ghost}} \,=\, 0.32(7) \;.
\en
We see that the values are respectively consistent with 
$13/22\approx 0.59$ and $9/44\approx 0.20$ within error bars
(but notice that there is a discrepancy of almost two standard 
deviations for $\gamma_{\mathrm{ghost}}$). 
In this case we have not succeeded in finding a value of $\omega$ 
describing the behaviors for $Z_3$ and $ \widetilde{Z}_3$ simultaneously.

\vskip 0.3cm 
In order to interpret the product of ghost and gluon form factors 
as the running coupling strength (see Subsection~\ref{sec:runcoup}), 
it is of great importance that the 
ghost-ghost-gluon-vertex renormalization constant $\tilde{Z}_1$ 
be finite in the continuum limit. 
For detecting the UV behavior of $\tilde{Z}_1$, let us investigate the 
product 
\be 
 Z_3(\beta, \mu)  \; \widetilde{Z}^2_3(\beta, \mu)  \; = \; 
 \frac{ \alpha_R(\mu^2) }{  \alpha_{0} (\Lambda _{UV}) } \; 
\widetilde{Z}^2_1(\beta, \mu) \; , 
\label{zprod} 
\en 
where (\ref{eq:alphaR}) was used. The left-hand side of the latter 
equation can be directly obtained from the numerical result 
for the renormalization constants $Z_3(\beta, \mu)$ and  
$ \widetilde{Z}_3(\beta, \mu)$. Note that for large UV cutoff, one 
finds 
\be 
\frac{1}{ \alpha_{0} (\Lambda _{UV}) } \; \propto  \; 
\ln \frac{ \Lambda ^2 _{UV} }{ \Lambda ^2 _{1-loop} } \; = \; 
- \, \ln ( \sigma a^2) \; + \; \mathrm{constant}, 
\en 
where the constant comprises cutoff (and therefore $\beta $) 
independent terms. 
The crucial point is that if the product $ Z_3(\beta, \mu)  \;
\widetilde{Z}^2_3(\beta, \mu)$ rises linearly with $-  \ln ( \sigma a^2) $ 
the additional factor $\widetilde{Z}^2_1(\beta, \mu)$ must be 
finite in the continuum limit 
[since the renormalized coupling $\alpha_R(\mu^2)$ is assumed finite]. 
Our numerical findings for $ Z_3(\beta, \mu)  \;
\widetilde{Z}^2_3(\beta, \mu)$ are also shown in Figures $\ref{fig:1}$ and
$\ref{fig:1B}$
(right panel). The data nicely support Taylor's findings, i.e.\ 
$\widetilde{Z}_1$ is cutoff- and therefore $\mu$-independent.

\subsection{The running coupling constant} 
\label{sec:ra}

\begin{figure}[t]
\vspace{-5.4cm}
\centerline{  
\epsfxsize=7cm 
\epsffile{alpha3l.eps}
\hspace{3mm}
\epsfxsize=7.5cm
\epsffile{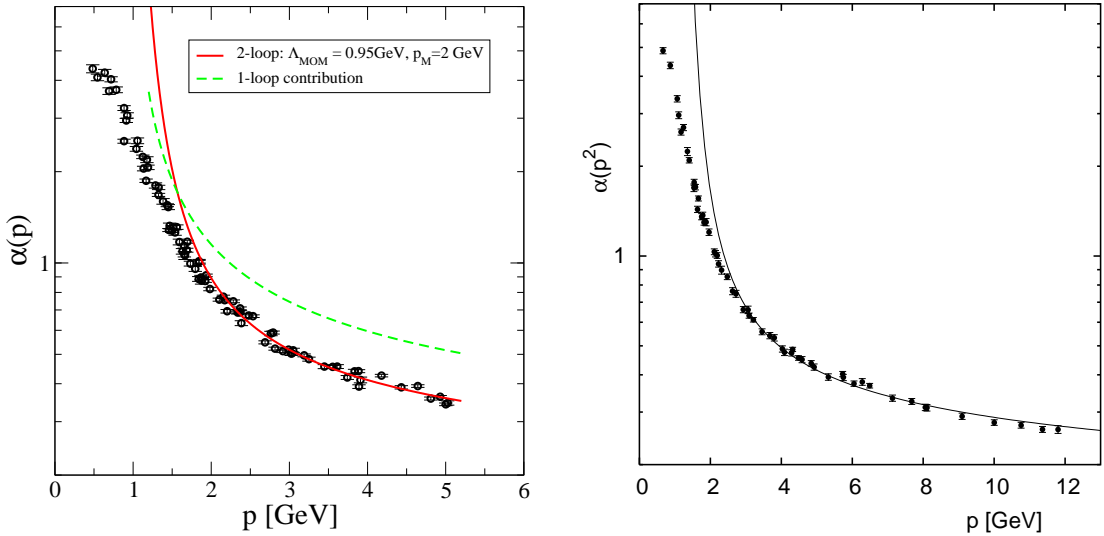}
           } 
\caption{ The running coupling in comparison with 
  the results from  perturbation theory:  T\"ubingen data (left panel) 
  and  S\~ao Carlos data (right panel). 
  The momentum cutoff is $p_M=2 \, $GeV in the former case and
  $p_M=2.5 \, $GeV in the latter. 
           }
\label{fig:33}
\end{figure}
Once it is established that $\widetilde{Z}_1$ is finite,
the momentum dependence of the 
running coupling constant can be simply derived from the product 
(\ref{eq:alphaRfin}) 
\be
\alpha_R(p^2) \; = \; 
\alpha_R(\mu^2) \; F_R(p^2,\mu ^2) \; J_R^2(p^2, \mu^2) \; . 
\en
The overall normalization factor can be obtained by comparing 
the lattice data with the well-known perturbative result, which 
is valid at high momentum. Here, we compare with the
2-loop expression, which is known to be independent of the 
renormalization prescription, i.e., 
\be
\alpha _{2-loop} \Bigl( x= p^2/\Lambda ^2_{2-loop} \Bigr) 
\;=\;
\frac{ 4 \pi }{\beta _0 \, \ln x } \Biggl\{
1 \; - \; \frac{ 2 \beta _1 }{ \beta ^ 2_0 } \frac{ \ln ( \ln x )
}{ \ln x } \Biggl\} 
\; , 
\label{eq:2-loop}
\en
with $\beta_0$ and $\beta_1$ given in eq.\ \reff{eq:b0b1}.
In order to obtain  $\Lambda _{2-loop}$ and to fix the overall factor, 
we fitted $ \alpha _R(\mu ) \,  F_R(p^2,\mu ^2) \; J_R^2(p^2, \mu^2) $ 
to the 2-loop running coupling $\alpha _{2-loop}(p)$ where 
only momenta $p \ge p_M$  were taken into account. Fitting parameters  
were $ \alpha _R(\mu )$ and the 2-loop perturbative scale 
$\Lambda _{2-loop} $. 
Starting from a very low value $p_M$ we fit these parameters while 
gradually increasing $p_M$. For small values of $p_M$, we observe that 
the functional form of (\ref{eq:2-loop}) tries to incorporate 
genuine non-perturbative effects by adjusting $\Lambda _{2-loop} $, thus, 
introducing a spurious $p_M$ dependence to $\Lambda _{2-loop}$.
However, a plateau is reached for the T\"ubingen data at
$p_M \approx 2 \, $GeV indicating that the data are well 
reproduced by the 2-loop formula in this regime. We find in this case that
\be 
\Lambda _{2-loop} \; = \; 0.95(15) \; \mathrm{GeV} \;.
\label{eq:mom}
\en 
For the S\~ao Carlos data we have cut the data at $p_M \approx 2.5\, $GeV,
which corresponds to a large drop in the $\chi^2/d.o.f.$ of the fit.
(This also corresponds to reaching a relatively good plateau for
$\Lambda _{2-loop}$ obtained from the fit.)
We obtain the value $\Lambda _{2-loop} \; = \; 1.2(1) \; \mathrm{GeV}$.
The two values are consistent within error bars.

Our final results for the running coupling constant are presented 
in Figure \ref{fig:33}. 
In the IR region, the two data sets show a clear departure from the
perturbative behavior and suggest a finite value $\alpha_c$ for the 
running coupling constant at zero momentum. We estimate $\alpha_c=5(1)$.
This value is in agreement with our previous fits for $\alpha_R(p^2)$
\cite{Bloch:2002we,Langfeld:2002dd}, and is consistent with the 
DSE result of Ref.~\cite{Bloch:2003yu}, $\alpha _c \approx 5.2$, using 
$\kappa = 0.5$ as input.


\subsection{Gluon and ghost form factors} 
\label{sec:factors}

\begin{figure}[t]
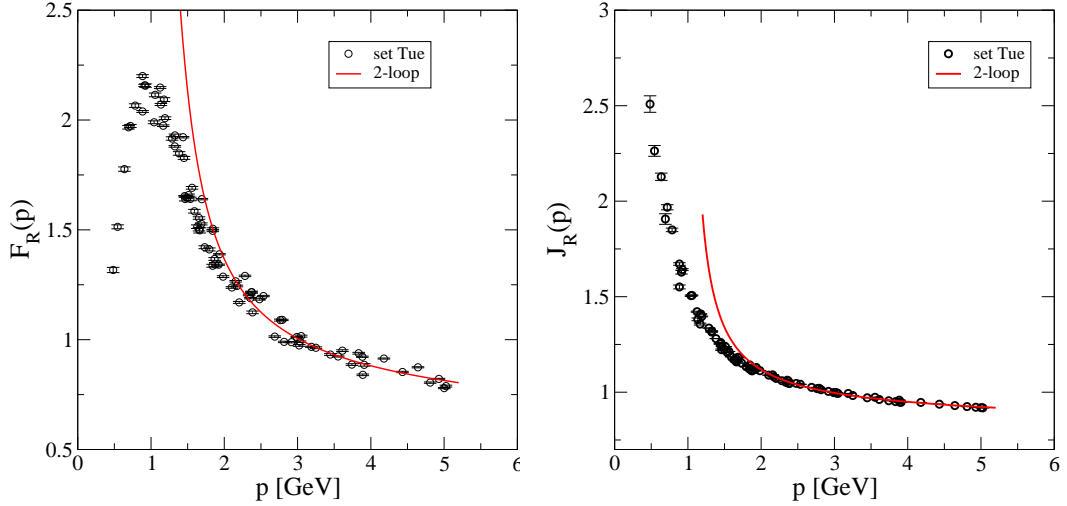

\centerline{
\epsfxsize=7cm
\epsffile{gluon2.eps}
\epsfxsize=7cm
\epsffile{ghost.eps}
}
\caption{The gluon form factor $F_R(p)$ and the ghost form factor $J_{R}(p)$ 
   as a function of the momentum transfer $p$ (T\"ubingen data).}
\label{fig:2}
\end{figure}
\begin{figure}
\begin{center}
\vskip -12.0cm
\includegraphics[width=14cm,angle=0]{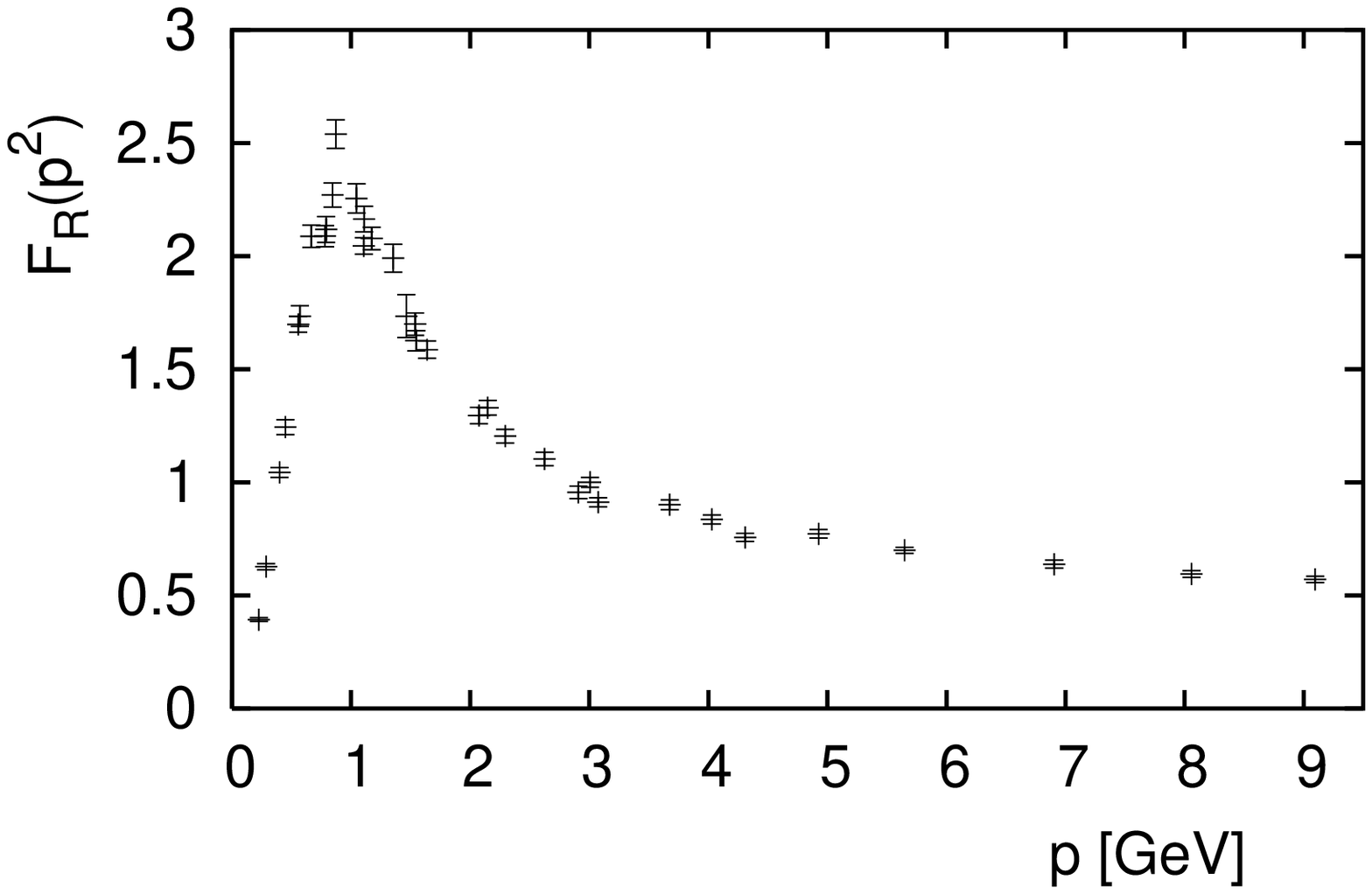}
\vskip -11.5cm
\hskip -0.4cm
\includegraphics[width=14.4cm,angle=0]{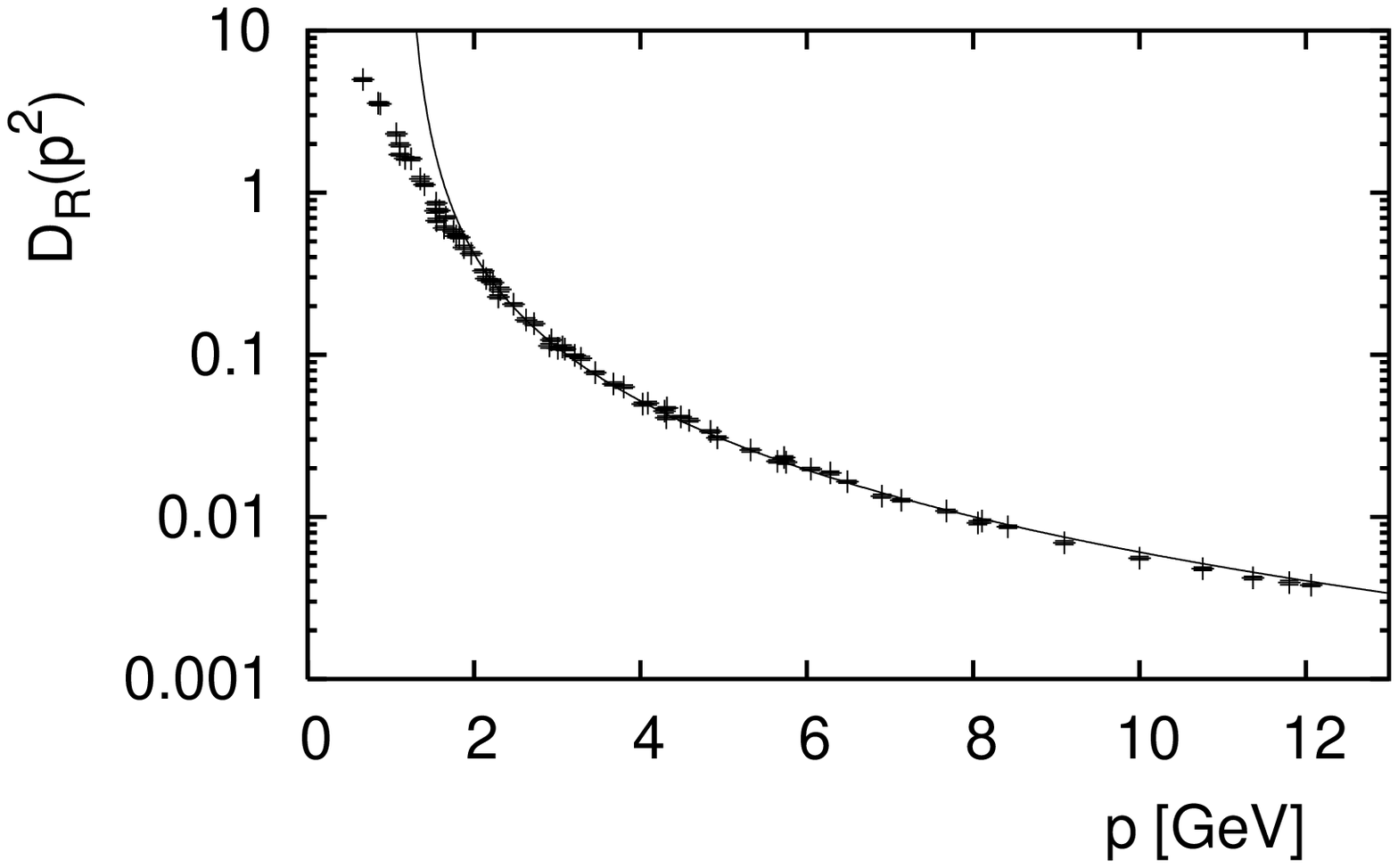}
\vskip 0.5cm
\caption{\label{fig:gluonplot} The gluon form factor
         $F_{R}(p)$ (above) and the
         gluon propagator $D_{R}(p)$ (below)
         as a function of the momentum transfer $p$
         (S\~ao Carlos data).
         Note the logarithmic scale in the second plot.
         Fit of gluon propagator using the analogue of
         eq.\ \reff{eq:form2loop} has been done for momenta $p \geq 2$ GeV
         and with $\Lambda = 1.2$ GeV.}
\end{center}
\end{figure}
\begin{figure}
\begin{center}
\vskip -12.0cm
\includegraphics[width=14cm,angle=0]{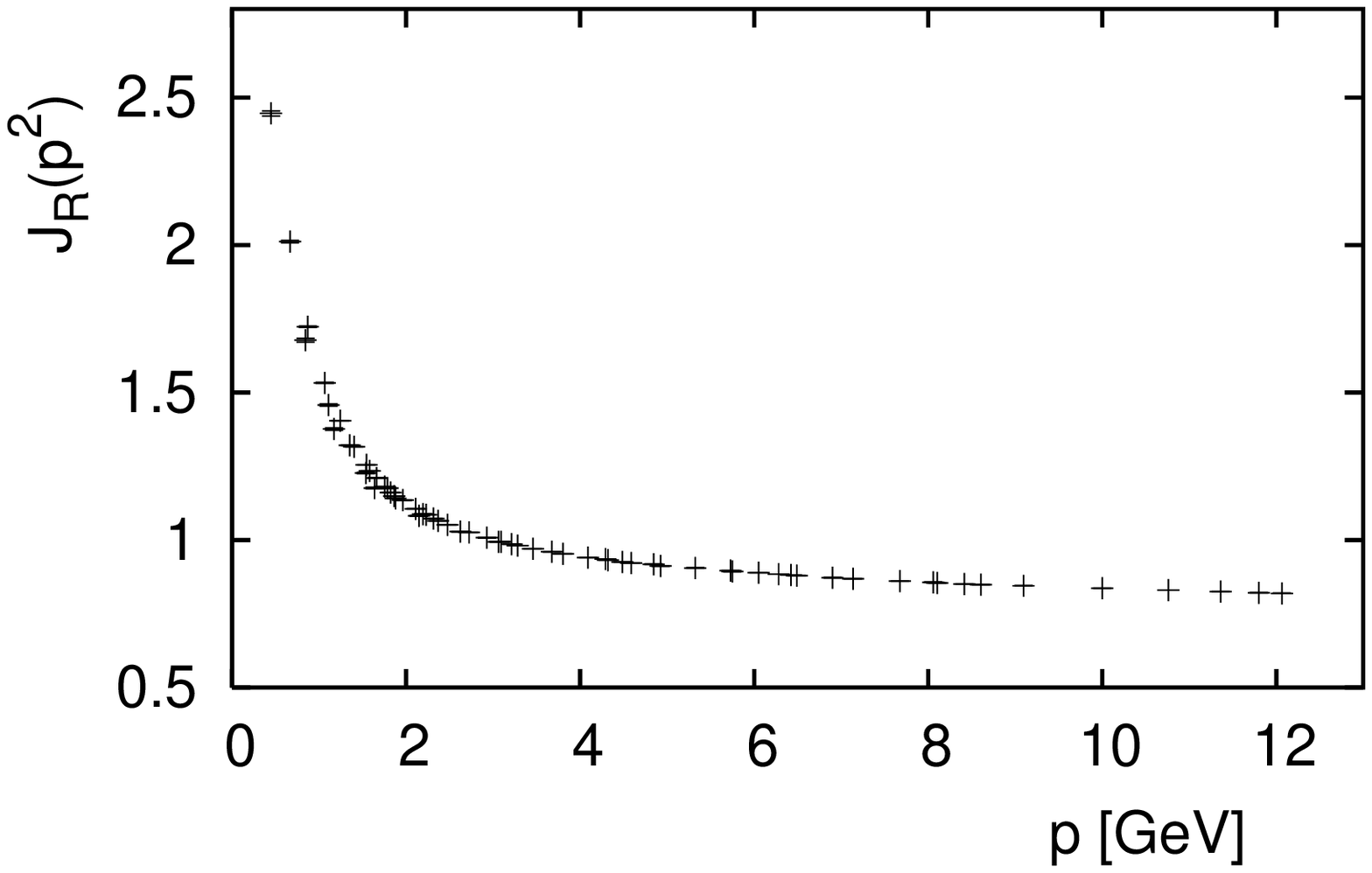}
\vskip -11.5cm
\hskip -0.4cm
\includegraphics[width=14.4cm,angle=0]{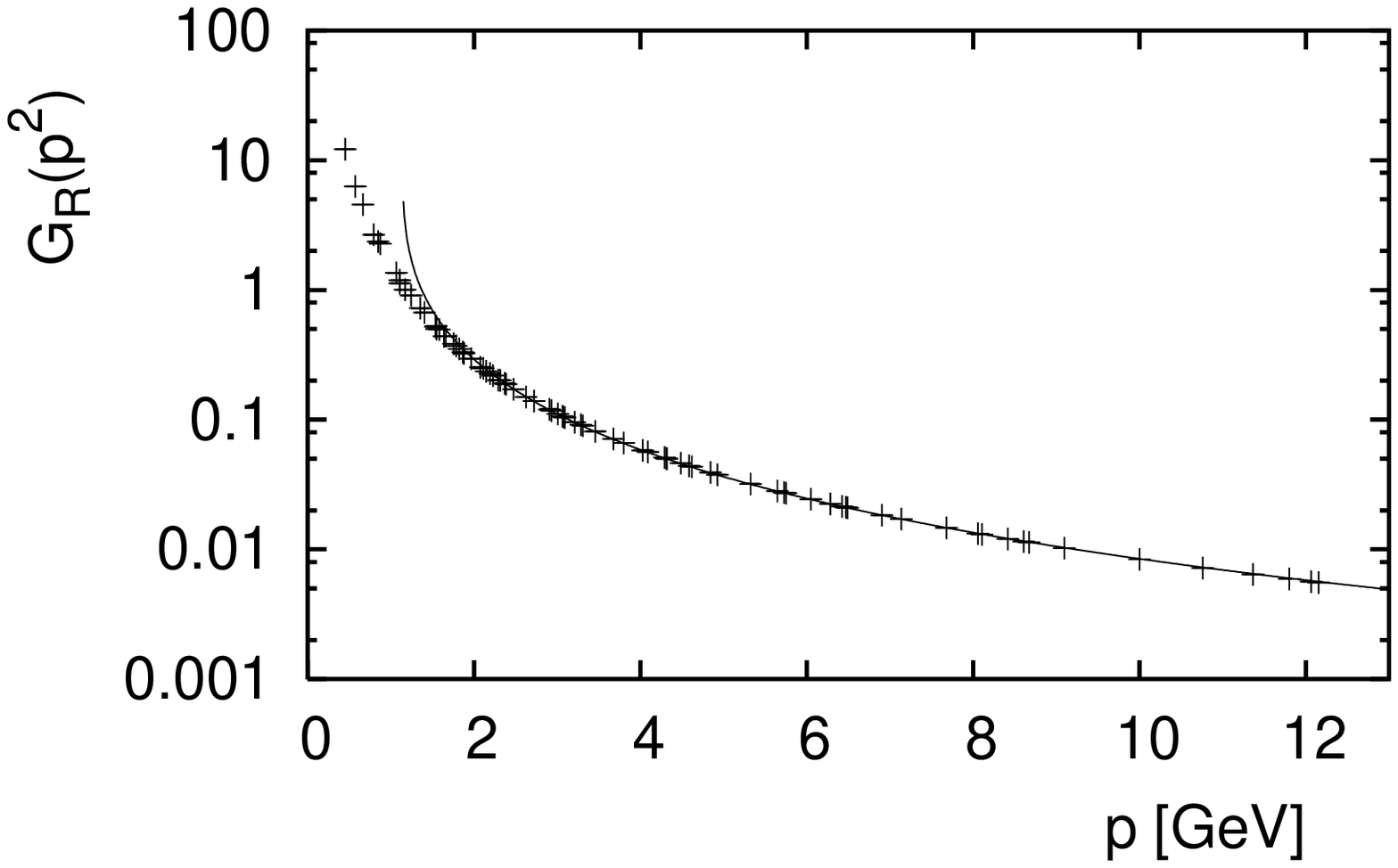}
\vskip 0.5cm
\caption{\label{fig:ghostplot} The ghost form factor
         $J_{R}(p)$ (above) and the
         ghost propagator $G_{R}(p)$ (below)
         as a function of the momentum transfer $p$
         (S\~ao Carlos data).
         Note the logarithmic scale in the second plot.
         Fit of ghost propagator using the analogue of
         eqs.\ \reff{eq:form2loop}
         has been done for momenta $p \geq 2$ GeV
         and with $\Lambda = 1.2$ GeV. }
\end{center}
\end{figure}

Figure \ref{fig:2} 
shows the T\"ubingen data for the gluon and the ghost form factors. 
The error bars comprise statistical errors only. It turns 
out that one observes an additional scattering of the data 
points which is not of statistical origin. This additional 
systematic noise is pronounced when simulated annealing is used 
for gauge fixing and it afflicts especially the small momentum range.
We attribute this error to the residual uncertainty of gauge fixing 
(Gribov noise). In particular, since the simulated annealing
is capable of hopping from one local minimum to the other, the
algorithm is sensitive to the large-scale structure of the minimizing
functional. This hopping then produces
a non-Gaussian noise which is underestimated when one uses
the standard Gaussian error propagation. For this reason,
we dropped the first three momentum points from the T\"ubingen
data sets. 

In Figs.\ \ref{fig:gluonplot} and \ref{fig:ghostplot} we report
the rescaled S\~ao Carlos
data for the renormalized gluon (respectively ghost)
form factor and the data for the corresponding 
propagators. We stress that in the gluon case finite-size
effects depend on whether we consider the full propagator or
the form factor. In fact, for the propagator these effects are
larger in the IR region, while for the form factor
the effects are larger in the UV
limit (due to the multiplication by $p^2$).
Thus, the ranges of momenta (for each $\beta$)
considered for the plots are different in the two cases.
Nevertheless, the matching factors obtained are in agreement.
The difference in finite-size effects between propagator and
form factor is less pronounced when considering the ghost propagator.

\vskip 0.3cm
At sufficiently high momentum $p \ge p_M$ (we found 
in Subsection \ref{sec:ra} that $p_M \approx 2 \, $GeV is an 
acceptable choice), the momentum dependence of
the renormalized form factors should be given by the
formula
\be
F_R (p^2 , \mu^2) , \; J_R (p^2 , \mu^2)
\; \approx \; d_2(\mu) \; \biggl[  \alpha _{2-loop} \Bigl(
\frac{p^2}{\Lambda ^2_{2-loop}} \Bigr) \biggr]^{\gamma} 
\; \biggl[ 1 \; + \; \bar{\gamma } \,  \alpha _{2-loop} \Bigl(
\frac{p^2}{\Lambda ^2_{2-loop}} \Bigr) \biggr]
\label{eq:form2loop}
\en
where $\gamma$ is the leading-order anomalous dimension of the gluon 
(respectively ghost) propagator, given by $\gamma = 13/22 $ 
(respectively $\gamma = 9/44 $).
The parameter $\bar{\gamma } $ stems from the next to leading order 
to the anomalous dimension and is scheme-dependent. At least 
in the $\overline{MS}$ scheme, this parameter is small. 
Furthermore, 
Equation (\ref{eq:form2loop}) can be derived from the renormalization-group 
equation using the 2-loop scaling functions $\beta (g_r)$ and 
$\gamma _A(g_R)$. Hence, (\ref{eq:form2loop}) originates from 
the resummation of an infinite set of 2-loop diagrams and, e.g., 
comprises the so-called ``leading logs''. 

\vskip 0.3cm 
Using the T\"ubingen data set, $p_M=2 \,$GeV and $\Lambda _{2-loop}=950 
\, $MeV, we fitted $\bar{\gamma }$ to the gluon and ghost data, 
respectively. We find that these parameters are indeed small, i.e., 
\be 
\bar{\gamma }_{\mathrm{gluon}} \; = \; - 0.036(18) \; , \hbo 
\bar{\gamma }_{\mathrm{ghost}} \; = \;   0.011(10) \; . 
\en 
Although the errors on these parameters are rather large, we find it 
encouraging that the parameters appear with opposite signs. 
In the case that the product of form factors $F_R(p^2) \,  J^2_R(p^2)$ 
is indeed renormalization group invariant, one would expect that 
\be 
\bar{\gamma }_{\mathrm{gluon}} \; + \; 2 \, \bar{\gamma }_{\mathrm{ghost}} 
\; = \; 0 \, . 
\en 
It is clear from the Figures \ref{fig:1} and \ref{fig:2} that 
the high momentum tail is well reproduced by (\ref{eq:form2loop}). 

\vskip 0.3cm 
For the S\~ao Carlos data we have found that a cut at $p_M=2 \,$GeV 
corresponds to a large drop in the $\chi^2/d.o.f.$ of the fits, both
for the gluon and for the ghost cases.
We start by fitting the leading-order term only 
(i.e.\ ignoring $\bar{\gamma }$). We get
\be
\Lambda_{2-loop}\;=\;1.19(4) \quad and \quad\;
\Lambda_{2-loop}\;=\;1.13(2)
\en
respectively from the fits of the gluon and of the ghost propagator. These
values are consistent with the result $\Lambda_{2-loop} = 1.2(1)$, obtained 
in the previous section. We then fix this value for $\Lambda_{2-loop}$ and
perform fits with $\bar{\gamma }$ as a free parameter. 
We obtain a good description of the data (see Figs.\ \ref{fig:gluonplot} and 
\ref{fig:ghostplot}), with values for $\bar{\gamma }$ even 
consistent with zero. 

As can be seen from our plots, the gluon form factor is suppressed
in the low momentum regime,
while the ghost form factor is divergent. Correspondingly, the ghost
propagator diverges faster than $1/p^2$
and the gluon propagator appears to be finite.
As mentioned in the Introduction, an IR-finite gluon propagator 
\cite{Cucchieri:1997fy}--\cite{Cucchieri:1997dx}
and an IR-divergent ghost form factor \cite{Suman:1995zg,Cucchieri:1997dx}
were obtained before by separate studies. 

A quantitative analysis of the IR behavior for the propagators --- including 
the evaluation of the exponent $\kappa$ mentioned in the Introduction --- was 
already presented in Refs.\ \cite{Bloch:2002we,Langfeld:2002dd}.
More thorough such analyses will be presented separately for the two sets 
of data in Refs.\ \cite{inprep_Tue,inprep_SC}.


\section{Conclusions} 
\label{conclusions}

For the first time, evidence from extensive lattice simulations 
is provided that the 
ghost-ghost-gluon-vertex renormalization constant $\tilde{Z}_1$ 
is indeed finite in continuum field theory (as found by Taylor using 
all orders perturbation theory). 
Also, our result is probably not affected by Gribov ambiguities, since
$\tilde{Z}_1$ is obtained using data in the UV limit.
It therefore appears that the Gribov ambiguities 
(in the lattice approach and the Faddeev-Popov Quantization)
do not afflict the renormalization of the vertex. 

Also, we performed a thorough study of the gluon and the 
ghost form factors. Our data favor the scenario of an IR finite 
(or even vanishing) gluon propagator while the ghost form factor
is singular in the IR limit. 

Finally, we have obtained the running coupling 
constant over a wide range of momenta using the data 
for gluon and ghost form factors. Our data are consistent with 
the existence of an IR fixed point $\alpha _c = 5(1)$. 
Note that this value is inside the interval given by the DSE
expression \reff{alpha_bounds}.

We stress that we compared our results for two slightly different
lattice formulations, obtaining consistent results in all cases considered.


\section*{Acknowledgments}

JCRB's contribution was funded by the Deutsche Forschungs\-gemeinschaft 
(Project no.\ SCHM 1342/3-1) and by the DFG Research Center "Mathematics for 
Key Technologies" (FZT 86) in Berlin. 
KL greatly acknowledges the stimulating atmosphere and the hospitality
of the Institute for Theoretical Physics, University 
of Karlsruhe where parts of the project were performed. 
The research of A.C.\ and T.M.\ is supported by FAPESP
(Project No.\ 00/05047-5). We thank C.~S.~Fischer for helpful 
comments on the manuscript. 



\end{document}